\documentclass[vecphys]{svmult}  
\usepackage[dvips]{graphicx}
\usepackage{amssymb}
\usepackage{makeidx}
\usepackage{multicol}
\usepackage[bottom]{footmisc}
\usepackage{cite}
\makeindex
\newcommand{\ket}[1]{\left| #1\right\rangle} 
\newcommand{\vev}[1]{\left\langle #1\right\rangle}

\newcommand{\GeV}{\; \mathrm{GeV}} 
 
\newcommand{\lapproxeq}{\lower .7ex\hbox{$\;\stackrel{\textstyle  
<}{\sim}\;$}} 
\newcommand{\gapproxeq}{\lower .7ex\hbox{$\;\stackrel{\textstyle  
>}{\sim}\;$}} 
\newcommand{\stackdown}[2]{\lower 1.4ex\hbox{$\;\stackrel{\textstyle{#1}}  
{\scriptstyle{#2}}\;$}}
\newcommand{\beq}{\begin{equation}} 
\newcommand{\eeq}{\end{equation}} 
\newcommand{\bea}{\begin{eqnarray}} 
\newcommand{\eea}{\end{eqnarray}}
\newcommand{\lsp}{\tilde{\chi}}

\makeatletter 
\def\slash{\@ifnextchar[{\fmsl@sh}{\fmsl@sh[0mu]}} 
\def\fmsl@sh[#1]#2{% 
  \mathchoice 
    {\@fmsl@sh\displaystyle{#1}{#2}}% 
    {\@fmsl@sh\textstyle{#1}{#2}}% 
    {\@fmsl@sh\scriptstyle{#1}{#2}}% 
    {\@fmsl@sh\scriptscriptstyle{#1}{#2}}} 
\def\@fmsl@sh#1#2#3{\m@th\ooalign{$\hfil#1\mkern#2/\hfil$\crcr$#1#3$}} 
\makeatother 
\begin{document}
\title*{LSP as a Candidate for Dark Matter}
\author{A.~B.~Lahanas\inst{1}}
\institute{University of Athens, Physics Department,  
Nuclear and Particle Physics Section, 
GR--15771  Athens, Greece
\texttt{alahanas@phys.uoa.gr}}
\maketitle
The most recent observations by the WMAP satellite provided us with 
data of unprecedented accuracy regarding the parameters describing the Standard Cosmological Model. The current matter-energy density of the Universe is close to its critical value of which 73 $\%$ is attributed to Dark Energy,  23 $\%$ to Cold Dark Matter and only 4 $\%$ is ordinary matter of baryonic nature. The origins of the Dark Energy (DE) and Dark Matter (DM)  constitute  the biggest challenge of Modern Astroparticle Physics. Particle theories, which will be tested in the next round experiments in large accelerators, such as the LHC, provide candidates for DM while at the same time can be consistent with the DE component. We give a pedagogical account on the DM problem  and the possibility that this has supersymmetric origin. 
%%
%%%%%%%%%%%%%%%%%%%%%%%%%% Paper body %%%%%%%%%%%%%%%%%%%%%%%%%
\section{Introduction}
The first evidence for Dark Matter (DM) stemmed from observations of clusters of Galaxies which are aggregates of a few hundred to a few thousand galaxies otherwise isolated in space. In 
$\; 1930 \;$ Smith and Zwicky examined two nearby clusters, Virgo and Coma, and found that the velocities of the galaxies making up the clusters were about ten times larger than they expected. This may be explained by assuming that there is more mass in the clusters which accelerates the galaxies to higher velocities. In $\; 1970 \;$ more reliable data, by observation of a larger number of clusters by Rubin, Freeman and Peebles, confirmed that the velocities of the galaxies are indeed different than one expects assuming that all matter comprising the galaxies is luminous. As you go to the edge of a spiral galaxy the amount of the light stars emit falls off and if all matter were luminous the rotational speed would fall off too. In fact 
from the distribution of luminous stars, the rotational velocity at a distance $r$ from the center of the galaxy turns out to be 
$\;v(r) \; \approx \; r^{-1/2}\;$ while observations showed instead that $\;v_{obs} (r) \; \approx \; const\;$. 
This cannot be explained unless there is some sort of invinsible matter or ``Dark", not interacting electromagnetically therefore, which participates however in the gravitational dynamics  
{\footnote{ Another explanation would be to assume that the gravitational force does not follow the simple inverse square law at galactic distances which although cannot be excluded it is rather ad-hoc lacking a firm theoretical foundation.}}.

Cosmologists usually measure the amount of mass and energy of the Universe in units of the critical density $\; \rho_c \;$ by defining the fraction $\; \Omega \equiv \rho / \rho_c \;$, with  $\; \rho\;$ is the mass-energy density of the Universe. When $\; \Omega > 1 \;$ Universe closes. Its value is accurately determined  by WMAP \cite{WMAP} due to the high precision measurements of the Cosmic Microwave Backround (CMB), $\; \Omega=1.02 \pm 0.02 \;$, confirming previous claims that our Universe is almost flat in concordance with inflation. After combining WMAP with other existing data and using the rescaled Hubble constant $\;h\;$ we have for 
the directly measured quantities $\; \Omega h^2 \;$, which are Hubble parameter independent, 
$\; \Omega_{matter} h^2 =0.134 \pm 0.006 \;$, of which a small amount 
$\; \Omega_{b} h^2=0.023 \pm 0.001 \;$ is of baryonic nature while the corresponding luminous mass density is smaller by an order of magnitude. 
The deficit $\Omega_{matter} h^2- \Omega_{b} h^2\;$  is attributed to Dark Matter whose value at the $2\sigma$ level lies within the range 
$$
0.094<\Omega_{DM} h^2<0.129 \quad .
$$
The value of the rescaled Hubble parameter is $\; h= 0.73 \pm 0.05 $ from which one can infer the values of $\; \Omega $'s. The conclusion is that 
our Universe is dominated by a large amount of energy  $\approx 73 \% $, of uknown origin, the so called Dark Energy (DE) and a large amount of mass $\approx 23 \% $, whose composition is also uknown the so called Dark Matter (DM). Only a small fraction, $\approx 4 \% $, consists of ordinary matter. Therefore $96 \%$ of our Universe is a completely mystery ! \\
%%%%%%%%%%%%%%%%%%%%%%%%%%%%%
%%%%%%%%%%%%%%%%%%%%%%%%%%%%%%%%%%%%%%%%%
Candidates for DM can be the neutrinos, axions, gravitinos or WIMPs or other more exotic particles 
such as cryptons, Kaluza-Klein excitations or branons existing in higher dimensional theories and so on.
The energy loss limit from SN 1987A put upper limits on axion masses, $\;m_a \leq 10^{-2} eV\;$, and on these grounds they are considered as non-thermal relics with very small mass. From the requirement that axions do not overclose the Universe lower limits on axion mass are imposed if one follows standard scenarios. The allowed mass window in standard considerations is very narrow leaving little room to believe that the axion can explain the DM of the Universe. For other more exotic as yet undiscovered candidates, proposed in higher dimensional gravity theories, already mentioned above, the situation is more involved and we do not put them under consideration in these lectures. 

Standard Model neutrinos and their antiparticles are existing particles and once believed that they could explain the Universe missing mass problem. However this possibility is rather ruled 
out in view of the latest data. If neutrinos are massless their density is  
$\; \Omega_{\nu}\approx 3.5 \times 10^{-5} \;$ and their contribution to the energy - matter density of the Universe is quite small.
However we now know that neutrinos are not completely massless and their relics are given by 
$\; \Omega_{\nu} h^2 = {\sum \; m_i}/{92.5} \;$. This assumes that their temperature today is $\; T_{\nu}=1.95 \; ^0 K = 1.7 \times 10^{-4} eV$. Therefore they could offer as  Hot Dark Matter (HDM) candidates. In general the possibility that the Universe is dominated by HDM seems to conflict with numerical simulations of structure formation. In fact relativistic matter streaming from overdense to underdense regions prevents structures from growing below the so called free-streaming scale \cite{colin}. In fact the combined results from WMAP and other data imply that 
$\; \Omega_{\nu} h^2 < 0.0076 \;$, providing also limits on neutrino masses,  and this is too low to account for the Universe's missing mass. 
Warm Dark Matter (WDM) neutrinos, or other low mass species, also seem unlikely. 
The reason is that star formation occurs relatively late in WDM models because small scale structure is suppressed. This is in conflict with the low-l CMB measurements by WMAP which indicate early re-ionization   (~at $z \approx 20 $) and therefore early star formation.{\footnote{ We have assumed that structure formation is responsible for re-ionization.}}
Actually  it is found that WDM candidates are incosistent with WMAP data for masses $m_X \leq 10 \; KeV$ \cite{yoshida} while masses larger than $ 100 \; KeV$ are almost indistinguishable from Cold Dark Matter. Cold Dark Matter is however  allowed and it is perhaps the most plausible possibility and candidates that may play this role naturally exist in some particle theories notably in Supersymmetric and Supergravity theories which are believed to be the low energy manifestation of String theories. 

%%%%%%%%%%%%%%%%%%%%%%%%%%%%%%%%%%%%%%%%%%%%%%%%%%%%
\section{The energy - matter content of the Universe}
The Universe is homogeneous and isotropic at supergalactic scales being therefore described by the 
Friedmann-Robertson-Walker (FRW)  geometry whose  metric is read from the line element 
\cite{weinberg,kolb}
\beq
d s^2\,=\, - c^2\;d t^2 \, + \, a^2(t)\, (  \frac{ d r^2}{1 \,-\,k r^2}  \, + \,
r^2 ( d \theta^2 \, + \, \sin^2 \theta d \phi^2 ) ) \; \; .
\eeq
%%%%%
In it  $a(t)$ is the cosmic scale factor and depending on $k$  we distinguish three types of
Universe with the following characteristics
\vspace{.5cm}
%%%%%%%%%%%%%%%%%%%
\begin{center}
\begin{tabular}{cccc}
\hline 
k \,\,  &\,\, Type of Universe \,\, & \,\, 3-d curvature \,\, &
\,\, Spatial volume \,\, \\ \hline \hline
1 \,\, & closed  & $k \, a^{-2} > 0$  & $2 \pi^{2} a^{3}$    \\  
0 \,\, & flat    &    0     & $ \infty $   \\ 
-1 \,\,& open    & $k \, a^{-2} < 0$ & $\infty$  \\ \hline
\end{tabular}
\end{center}
%%%%%%%%%%%%%%%%%%%%%%%%%%%%%%%%%%%%%%%%%%%%%%%%

The expansion rate of the Universe is defined by
\bea
H \,=\, \frac{\dot{a}}{a} \,\,. \label{rate}
\eea
Its value today, the wellknown ``Hubble constant", is denoted by $H_0$ and can be 
written as $ H_0 =100\; h_0 \;Km/sec/Mpc \;$ 
{\footnote{1 pc $\approx$ 3.26 light years.}}
by defining the 
dimensionless quantity $h_0$ which is usually called ``rescaled Hubble's constant". Its value
is experimentally known with a fairly good accuracy by the  WMAP data, $h_0\;=\;0.72 \pm 0.05$.
The matter - energy density of the Universe is related to the expansion rate by 
\beq
\varrho \;=\;  \frac{3}{8 \pi G_N } \;( \; H^2 + \frac{k}{a^2} \;)
\; ,  \label{ex3}
\eeq
%%%
while the ``critical density" $\; \varrho_c \;$, is defined by,
\beq
{\varrho}_c \;=\;  \frac{3}{8 \pi G_N } \;H^2_0   \,\, . 
\label{crit}
\eeq
%%%
Its value is 
$ {\rho_c} \;=\; 1.88 \times 10^{-29} \;h_0^2 \;gr /cm^3 \;=\; 8.1 \times 10^{-47} \;h_0^2 \; GeV^4 $. By (\ref{ex3}) it is seen that 
if the value of the matter-energy density today is $ {\varrho}_0$ then
depending on whether this is larger, smaller or equal
to the critical density our Universe is close, open or flat respectively,  
\vspace{.5cm}
%%%%%%%%%%%%%%%%%%%
\begin{center}
\begin{tabular}{lll}
\hline \\
$ { \rho}_0 > {\rho_c}   $ &\,\,\,   $ k > 0 $ &\,\,\, closed Universe  \\
$ { \rho}_0 = {\rho_c}   $ \,\,\, $\Longrightarrow$ &\,\,\,   $ k = 0 $ &\,\,\, flat Universe   \\
$ { \rho}_0 < {\rho_c}   $ &\,\,\,   $ k < 0 $ &\,\,\, open Universe   \\ \\
\hline
\end{tabular}  
\end{center}
%%%%%%%%%%%%%%%%%%%%%%%%%%%%%%%%%%%%%%%%%%%%%%%%
Recent observations by WMAP and other sources point towards a flat Universe. 

The total matter-energy density today can be expressed as fraction of the
critical density by defining the ratio 
%%%%%%%%%%%%%%%%%%%%%%%%%%%%%%%%%%
\beq
\Omega \;=\; \frac{\varrho_0}{\rho_c}  \; .\label{omega}
\eeq
%%%%%%%%%%%%%%%%%%%%%%%%%%%%%%%%%
This can be written as a sum 
%%%%%%%%%%%%%%%%%%%%%%%%%%%%%%%%%%
\beq
\Omega  \;=\; \sum_i \;\Omega_i 
\eeq
%%%%%%%%%%%%%%%%%%%%%%%%%%%%%%%%%
with $\;\Omega_i \;$ denoting the contribution of each particle species, 
including the contribution of the radiation $\; \Omega_{radiation}$
and that of the cosmological constant $\; \Omega_{\Lambda}$. Note that $\Omega$'s as defined above refer to today's values. 
%%%%%%%%%%%%%%%%%%%%%%%%%%%%%%%%%%

From observations of distant Galaxies Hubble (1929) drew the conclusion that 
%%%%%%%%%%%%%%
\beq
d_L \; H_0 \;=\; c \;z \quad \label{hubble}
\eeq
%%%%%%%%%%%% 
In (\ref{hubble}) $d_L$ is the luminosity distance and $z$ the Doppler shift in the wavelength of the emitted radiation, $\; 1+z = \lambda_{observer}/ \lambda_{source}\;$. 
For a moving source with velocity $v$, which radiates electromagnetic radiation, $v=c z$ and therefore this law states that Galaxies recede with velocities given by 
%%%%%%%%%%%
$$
v_r \;=\; d_L H_0  \; .
$$
%%%%%%%%%%%
This is perhaps one of the most spectacular discoveries in the history of Astrophysics. 
{\emph{Our Universe is not static but it expands}}. FRW geometry is actually encompassing Hubble's law since the distance $d$ is proportional to the cosmic scale factor $a$ and hence the receding velocities are $v=H d$. \\
The luminosity distance, which is not actually the true geometric distance, between emitter and receiver, is defined by 
%%%%%%%%%%%%%%%%%%
\beq
%\mathrm{Flux}\equiv 
{\cal F} \;=\; \frac{\cal{L}}{4 \pi d_L^2} , \label{flux}
\eeq
%%%%%%%%%%%%%%%%%%%
where the flux $\cal F$ is the energy received in the apparatus per unit area, per unit time and $\cal{L}$ is the ``absolute luminosity".  $d_L$ would be the actual geometric distance to the Galaxy if our Universe were static. In an expanding  Universe $d_L$ is given by \cite{weinberg,kolb}
%%%%%%%%%%%%%%%%%%%%%%%%%%%%%%%%%%%%%%%%%%%%%%%
\bea
d_L\;=\; a(t_0) \; r_1(t_1) \; {(1+z)} \; . \label{lumdis}
\eea
%%%%%%%%%%%%%%%%%%%%%%%%%%%%%%%%%%%%%%%%%%%%%%%
In this  $t_0$ is observer's time and $t_1$ is the time the light signal was emitted from the Galaxy.  $r_1$ is given by 
$\int_0^{r_1} \frac{dr}{\sqrt{1-k r^2}} \;=\; \int_{t_{1}}^{t_0} \; \frac{d{t^{\prime}}}{a(t^{\prime})} \;$. 
The  $1+z$ in (\ref{lumdis})  effectively reduces the absolute luminosity $\cal{L}$ in 
( \ref{flux} ) by a factor ${(1+z)}^{-2}$. One power of ${(1+z)}^{-1}$ accounts for the fact that photons are redshifted  and their energy  $E^{\prime}$ when they reach the observer is $1+z$ times smaller than their energy when the were emitted. The other ${(1+z)}^{-1}$ power duly takes care of the fact that the number of photons per unit length of the beam drops as the light beam increases with expansion and thus the observer's apparatus receives less photons per unit time. Due to these two effects the observed luminosity is $\; {\cal L}_o \;=\;{\cal L}/ {(1+z)}^{-2}$ and (\ref{flux}) could have been equally well expressed as $\; {\cal F}= {{\cal L}_0}/{ 4 \pi d_G^2} \;$ with $\;d_G = a(t_0) r_1(t_1)\;$ the geometric distance to the Galaxy.  

The cosmological redshift is defined by 
%%%%%%%%%%%
\beq
 1+z = \frac{\lambda(t_0)}{ \lambda(t)}\; \label{zzz1} \; ,
\eeq
%%%%%%%%%%%%%%%%%%
where $\;{\lambda(t_0)}, { \lambda(t)}\;$ are respectively the wavelengths of the radiation 
today $t_0$ and at the time of emission $t$. If a signal is emitted from a source at $t$ and it is received at $t_0$ it is easy to show that 
%%%%%%%%%%%
\beq
  \frac{\lambda(t_0)}{ \lambda(t)}\;=\; \frac{a(t_0)}{ a(t)} \; 
\eeq
%%%%%%%%%%%%%%%%%%
which has the meaning that wavelengths also scale with the cosmic scale factor like distances. Using this we get from (\ref{zzz1})
%%%%%%%%%%%
\beq
 1+z = \frac{a(t_0)}{ a(t)} \; . \label{zzz2}
\eeq
%%%%%%%%%%%%%%%%%%
This relation between redshists and cosmic scale factor can be used to express the time $t$ as function of the redshift $z$. Today $z=0$ while $z=\infty$ at the beginning, $t=0$. By using  
$\; H={\dot a} / a \;$ it follows that 
%%%%%%%%%%%%%%%
\beq
t(z)\;=\;\int_z^{\infty} \; \frac{dz}{H\;(1+z)} \;, \label{tz}
\eeq
%%%%%%%%%%%%%%%
where $H$ as function of $z$ is given by 
%%%%%%%%%%%%%%%%%
\beq
H(z)\;=\;H_0 \;{[\; \sum_i\; \Omega_i \; {(1+z)}^{3(1+w_i)} +  
(\;1 - \sum_i\; \Omega_i \; ) {(1+z)}^{2} \;]}^{1/2}  \; .  
\eeq
%%%%%%%%%%%%%%%%%
If the redshift of some astronomical source is known then from (\ref{zzz2}) we can estimate the relative size of the Universe when light was emitted from it. For instance quasars are characterized by $z=5.82$ and from (\ref{zzz2}) we conclude that our Universe was $6.82$ times smaller when light was emitted from these objects. 

\begin{figure}[t]
\centering
\includegraphics[scale=0.65]{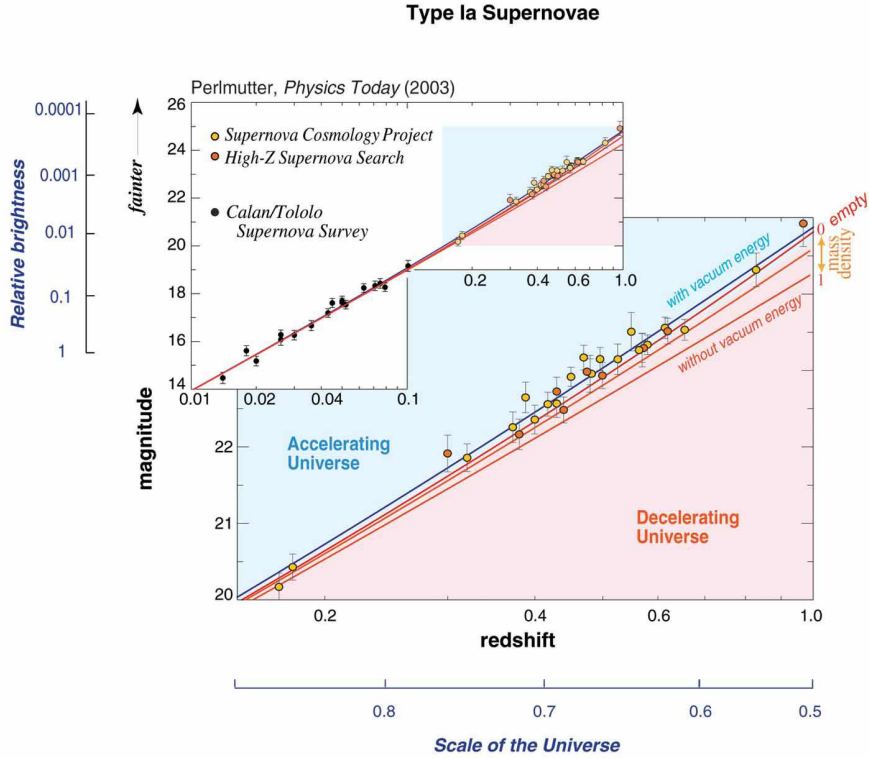} 
\caption[]{
The Hubble diagram for the high-z SNIa analyzed by the SCP and HZSS collaborations 
\cite{astrodata}.}
\label{SCP}
\end{figure}

The luminosity distance as function of the redshift can be written as 
%%%%%%%%%%%%%%%%%
\beq
d_L\;=\;c \;{H_0}^{-1}\; {|\Omega_k|}^{-1/2} (1+z) \;
sinn[\;{|\Omega_k|}^{-1/2}  H_0 \; \int_0^z \frac{dz}{H} \;] \; .  \label{dzzz}
\eeq
%%%%%%%%%%%%%%%%%
where the function $ sinn(\chi)$ is $\; sin(\chi), \chi $  or $ sinh (\chi) \;$ for 
$k=1, 0, -1 $ respectively. 
Assuming only contribution of matter and cosmological constant $\; \Omega_M, \Omega_{\Lambda} \;$ this can be expanded in powers of $z$ with the result
%%%%%%%%%%%%%%%%%
\beq
d_L\;=\;c {H_0}^{-1}\; [\; z \;+\; 
\frac{z^2}{2} \;(\;1+\Omega_{\Lambda}- \frac{\Omega_M}{2} \;) \;+ ... \;] \; . 
\label{dexpand}
\eeq
%%%%%%%%%%%%%%%%%
The ellipsis in (\ref{dexpand}) stand for terms of order $\sim z^3$ and higher. 
The deceleration parameter $q_0$ is defined by $\;q_0 \equiv - {\ddot a_0} a_0 / {a_0}^2\;$ and assuming existence of only matter and cosmological constant it is given by 
%%%%%%%%%%%%%%%%
\beq
q_0\;=\; - \Omega_{\Lambda}+\frac{\Omega_M}{2} \;. \label{decel}
\eeq
%%%%%%%%%%%%%
On account of it one observes that the coefficient of the $z^2/2$ term in (\ref{dexpand}) is exactly equal to  $\; 1-q_0 \;$. 
In (\ref{dexpand}) we have kept terms up to $\sim z^2$ which is necessary for the study of distant supernovae, as is the case with the SNIa, which are characterized by redshifts in the range $z = 0.16-0.62$. Observations showed that these objects appear fainter and hence at larger distances than expected. 
The analyses by SCP and HZSS collaborations \cite{astrodata} showed that the data are consistent with theory if one assumes the existence of a non-vanishing contribution $\Omega_{\Lambda}$ from   the cosmological constant which is correlated to $\Omega_M $. The value of $\Omega_{\Lambda}$ turns out to be about $ 70 \% $ showing in a spectacular manner that the bulk of the matter-energy density of the Universe is void~! From the values of $\Omega_{\Lambda,M}$ one can see that the deceleration parameter $q_0$ is negative or same the Universe is accelerating. The relevant Hubble diagram is shown in figure \ref{SCP}. Five years after the direct evidence for the existence of a nonvanishing cosmological constant which accelerates the Universe WMAP precision data confirmed previous analyses leaving little room for doubt that the bulk of the matter-energy of the Universe is vacuum energy while the majority of the matter is invinsible or Dark Matter \cite{WMAP}.

%%%%%%%%%%%%%%%%%%%%%%%%%%%%%%%%%%%%%%%%  
%%%%%%%%%%%%%%%%%%%%%%%%%%%%%%%%%%%%%%%%%
All discussion so far assumes that our space time is four-dimensional. The picture changes drastically if one assumes that there exist extra dimensions (ED)  that are invisible at present energies. These ideas are very popular nowadays in particle physics community and they are inevitable in string theories which require the existence of extra dimensions. Especially the last five years there is a lot of activity in the direction of the Brane World physics stirring new interest to the field with a variety of theoretical predictions that are waiting for their experimental confirmation ( or rejection ! ) in the next generation high energy accelerators. In such ED scenaria not only particle physics but also Cosmology is affected and the Standard Cosmological Model has to be reconsidered. For instance in a Universe confined to a 3-brane within a larger spacetime does not lead to Friedmann equations of the Standard Cosmology. In fact the Hubble expansion rate 
$\; H \equiv {\dot a}/{a} \;$ is not proportional to the density $\varrho$ but rather  to  $\varrho^2$ \cite{binetruy}. Assuming the bulk contribution is negligible at Nucleosynthesis one has 
$\; a(t) \sim t^{1/4} \;$. Moreover  $\; H \sim T^4 \;$ unlike $\; H \sim T^2 \;$ of the standard Cosmology. Therefore cooling is much slower and the freezing temperature $\;T_D\;$ of the neutron to proton ratio changes from about $0.8\; MeV $ to about $3 \;MeV$. Values of  $\;T_D\ \sim 2 \;MeV \;$ are then required by Big Bang Nucleosynthesis (BBN) and since $g^{-1/8}_{*} \; M_{(5)} \simeq 3 \; MeV$, with $M_{(5)}$ the scale characterizing the 5-dimensional Gravity, this relation cannot be satisfied. Actually more dimensions are required. Thus it appears that although ED seems a quite intriguing idea new parameters enter the game, at least as far as Cosmology is concerned, which may upset the successes of the Standard Cosmological Model. Within the context of these ED  scenaria various attempts to find a reasonable expanation for DE and DM have been put forward. In these lectures we shall be more conservative and pursue the idea that DM are non - Standard Model (SM) particles living in the standard four space-time dimensions. The issue of extra dimension will be covered by other speakers in this school.

%%%%%%%%%%%%%%%%%%%%%%%%%%%%%%%%%%
\section{The Thermal Universe} 
%\subsubsection{The Thermal Distributions of Particles}
At early times the ingredients of our Universe were the Standard Model particles, and perhaps additional as yet undiscovered particle species, being in a state of hot plasma in thermodynamic and chemical  equilibrium. At a given temperature $T$ the energy density of a particular particle species of mass $m$ is given by 
%%%%%%%%%
\begin{eqnarray}
\rho\;&=&\;\frac{ \pi^2}{30} \;{( kT )}^4 \;g(T) \;  . \label{density2} 
\label{pressure2} 
\end{eqnarray}
%%%%%%%%% 
The functions $g(T)$ can not be expressed in a closed form. However for temperatures, 
$ kT >> m  $, the particle is almost relativistic and $g(T)$ is just a constant given by 
%%%%%%%%%
\begin{eqnarray}
g = \;g_s \; {N^{\prime }}_{B,F}    \label{entropy22}  \;\;.
\end{eqnarray}
%%%%%%%%% 
The subscripts $B$ or $F$ stand for a boson or a fermion respectively,  
${N^{\prime }}_B =1$, ${N^{\prime }}_F =\frac{7}{8}$ and $g_s$ is the number of spin degrees of freedom. 

In the opposite limit, $ kT << m  $, the functions $g(T)$ approaches its well-known Boltzmann expression and the energy density is 
%%%%%%%%%
\begin{eqnarray}
\rho \;&\simeq & \; 
 g_s \;m\: { \left( {\frac{mkT}{2 \pi  }}\right)}^{\frac{3}{2}} \; \exp \;
{\left( -\frac{m }{kT} \right) } . \label{bol1} 
\end{eqnarray}
%%%%%%%%% 
In this limit $p <<  \rho$, that is, we deal with a pressureless gas or ``dust". 
%%%%%%%%%%%%%%%%%%%%
%\subsection{The Energy Densities of Photons and Neutrinos}

Especially for the photons and neutrinos the situation is as follows: \\
{ \bf { Photons :} }
The contribution of the photons to the total matter-energy density can be read from  
(\ref{density2}) with $g(T)=2$ since the photon is a massless spin-1 particle, having therefore two helicity degrees of freedom.  
Thus employing the fact that today $T_\gamma \simeq 2.7 \; \;^0K$, 
one has, 
%%%%%%%%%%
\beq
\Omega_{\gamma}\;\equiv \;\frac{\rho_{\gamma}}{\rho_c}\;\simeq \; {2.5}\;\times \; 10^{-5} / {h_0^2}   \;\;. \label{gamma}
\eeq
%%%%%%%%%%%%%
With $h_0 \approx 0.7$ this yields $\; \Omega_{\gamma}\simeq 5.2 \times 10^{-5}$. That is, the energy density carried by
photons is five orders of magnitude smaller than the critical density.
%%%%%%%%%%%%%%%%%%%%%%%%%
\\
{ \bf { Neutrinos :} }
If neutrinos or antineutrinos are massless then their energy density, for each species, is given by
%%%%
$$
\rho_{\nu } \;=\; \frac{7 \pi^2}{240}\;{(kT_{\nu})}^{4} \,\,.
$$
In deriving this we used the fact that neutrinos and antinieutrinos are massless and they are each characterized by one helicity state ( see Eq. (\ref{density2})). Their total contribution to $\Omega$ can be weighted relative to $\Omega_{\gamma}$ and is given by%%%%%%%%%%%%%%%%%%%%%%%%%%
%%%%%%%%%
\beq
\Omega_{\nu } \;=\; N_{\nu} \;\frac{7}{16} \; {\left( \frac{T_{\nu}}{T_{\gamma}} \right)}^{4} \;
\Omega_{\gamma } \; , \label{neutr}
\eeq
%%%%%%%
where $N_{\nu}$ counts the total number of neutrino and antineutrino species, i.e. $N_{\nu}=6$ 
in the Standard Model. One should note that the neutrinos' temperature $T_{\nu}$ is different from that of photons $T_{\gamma}$ and hence the subscripts in 
the equation above. The reason is that neutrinos decouple from the cosmic soup when the Universe's 
temperature is $k T_D \simeq \mathrm{MeV}$ and then they expand freely. At a lower temperature $k T \simeq 2 m_e$ photons reheat, due to the annihilation process $e^{+} e^{-} \; \rightarrow \; 2 \gamma $, and from entropy  
conservation it follows that $ T_{\nu} / T_{\gamma} \;=\;{\left( \frac{4}{11} \right)}^{1/3} $ after reheating. Using this we have from (\ref{neutr}), 
%%%%%%%%%%%%%%%%%%%
\beq
\Omega_{\nu } \; \simeq\;0.115 \; N_{\nu} \; \Omega_{\gamma} \; ,  \label{nus}
\eeq
 %%%%
from which we deduce that the energy density of massless neutrinos is also much smaller than the critical density of the Universe. 

\begin{figure}[t] 
\begin{center}
\includegraphics[width=7cm]{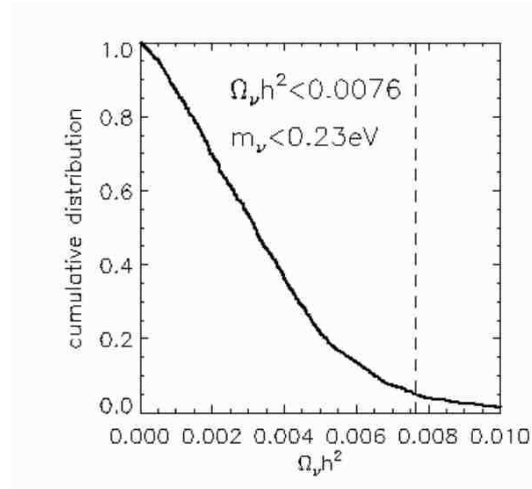}
\end{center}
\caption{{WMAP bounds on Neutrino masses and their relic density, \cite{WMAP}.}}
\label{neutra}
\end{figure}

The situation changes for massive neutrinos however. If we assume  a massive neutrino ( or antineutrino ), of mass $m_{\nu}$, then its energy density, at temperatures larger than its mass, is ( see (\ref{bol1}) ),  $\rho_{\nu} \simeq n_{\nu } \;m_{\nu} $ from which we have  
%%%%%%%%%
\beq
\Omega_{\nu} \; \simeq \; 0.0053 \; \frac{\sum_{\nu}  \left( m_{\nu}/ \mathrm{eV} \right) }{h_0^2} \; ,
\eeq
%%%%%%%%%
with $m_{\nu} $ in eV. In this expression the sum runs over all neutrino and antineutrino species.  Employing a value $h_0 \simeq 0.7$
this yields $\;\Omega_{\nu } \; \simeq \; {\sum_{\nu} m_{\nu} }/({92.5} \; \mathrm{eV}) \;$. 
The best evidence for neutrino masses comes from the {\bf{SuperKamionkande}} experiment 
\cite{kamioka} which detected 
oscillations $ \nu_{\mu}\; \rightarrow \; \nu_{\tau}  $. This result indicates a mass difference $m_{\nu_{\tau}}- m_{\nu_{\mu}} \simeq 0.05 \; \mathrm{eV} $ which points towards small neutrino masses of the order of  a few  $\mathrm{eV}$.  However experimental data from neutrino physics do not put an upper limit to neutrino masses able to exclude neutrinos as Dark Matter candidates. It is the WMAP data in combination with data from other sources that almost exclude the possibility that neutrinos can be either Hot or Warm Dark Matter as already mentioned in the introduction. In fact the limit  $\; \Omega_{\nu} h^2 < 0.0076 \;$ is extracted, from which an upper limit on $\;{\sum_{\nu} m_{\nu} }\;$ can be derived as is seen in figure \ref{neutra}. 
Therefore neutrinos contribute very little $\; \sim 10^{-2}\;$, or less, to $\; \Omega \;$, as photons do, and hence unable to account for the observed amount of Dark Matter. 
%%%%%%%%%%%%%%%%%%%%%%%%%%
%%%%%%%%%%%%%%%%%%%%%%%

In the evolution of the Universe one distinguishes two main epochs during which radiation and matter dominated respectively. 
\\
{ {\bf { Radiation Dominance :} }}
During radiation dominance, the energy density was dominated by 
relativistic particles whose thermal properies, apart from their spin content, is like that of photons. In this era $p\;=\;{\rho}/{3}$ and from energy conservation it follows that 
$\;d (\rho \; a^4 ) / dt= 0\;$ which is immediately solved to yield 
$\; \rho \simeq 1 / a^4\;$. Then from (\ref{ex3}) the density term dominates over $k/a^2$ and we have an equation for the cosmic scale factor
which can be solved to yield $\; a \simeq t^{1/2}\;$. From these two we get
%%%%%
\bea
\rho\;=\; \frac{3}{32 \;\pi\;G_N}\; t^{-2}  \,\,, \label{rhot}
\eea
%%%%%%%
which on account of (\ref{density2}) yields, 
%%%%%%%%%
\bea
T\;=\;{\left( \frac{16\;\pi^3\;G_N}{45} g(T) \right) }^{-1/4} \;t^{-1/2} \,\,. \label{TT}
\eea
%%%%%%%%%%%
Putting it all together, during the radiation era, the cosmic scale factor, the energy density and the temperature behave as 
\bea
a \sim t^{1/2} \, , \,\rho \sim t^{-2} \, , \, T \sim t^{-1/2} \; ,
\eea
%%%%%%%%
and the expansion rate is, 
%%%%%%%
\bea
H\;=\;1/2t\;=\; 1.66\;\sqrt{\;g(T)} \; \; \frac{ {(k T)}^2}{M_{Planck}} \,\,. \label{HH}
\eea
%%%%%%%%%%%%%%%%%%%%
The content of $g(T)$ entering (\ref{TT}) and (\ref{HH}) is different at different 
times. If a massive particle with mass $m_i$ is in equilibrium with the cosmic soup, then for 
temperatures $k T >> m_i $ it is relativistic and contributes either $7 / 8 \;g_s$ or $g_s$, depending on whether 
it is a boson or a fermion, see (\ref{entropy22}). As the Universe expands, its temperature drops and eventually 
reaches a temperature for which $k T << m_i$. Therefore, its contribution is 
negligible because of Boltzmann suppression $\exp \; (-{m_i}/{kT})$, see (\ref{bol1}), and  can be ignored. In table~\ref{tab1} the values of $g(T)$ at different temperatures are shown along with the radiation content at the  temperature shown on the right. For temperatures above the top mass,  $m_t \simeq 175.0 \; GeV$, the contributions of additional non Standard Model (SM) particles, if they exist, should be added. The contribution of the SM Higgs boson adds $+1$ to 
$g(T)$ for $k T > m_{Higgs}$ and is not shown. The LEP experimental limit put on the 
Higgs mass is $m_{Higgs} > 114 \; GeV$, \cite{higgsLEP}. We should remark that in table~\ref{tab1}, the effects of the neutrino decoupling and the photon reheat have not been counted for.
%%%%%%%%%%%%%%%%%%%%%%%%%%%%%%%%%%%TABLE1
\begin{table}
\caption{Active degrees of freedom and their contribution to g(T) for the Standard Model Particles}
\centering
\setlength\tabcolsep{15pt}
\begin{tabular}{lll}
\hline\noalign{\smallskip}
   k T  & Content of radiation &  $g(T)$ \\  
%\noalign{\smallskip}
\noalign{\smallskip}\hline\noalign{\smallskip}
 $< m_e$ & $ \gamma + 3 \times ( \nu + \bar{\nu}) $   &  29 /4 \\
 $m_e - m_\mu$ & $\cdots + e^{+} , e^{-}   $          &  43/4 \\
 $m_\mu - m_\pi $& $\cdots + \mu^{+} , \mu^{-}$       &   57/4 \\
 $m_\pi - \Lambda_c $ & $ \cdots + \pi^{+},\pi^{-} , \pi^{0} $ & 69/4 \\
 $\Lambda_c - m_s$ & $ \cdots + u, \bar{u}, d, \bar{d}, gluons $ & 205/4 \\
 $m_s - m_c$ &  $ \cdots + s, \bar{s} $ & 247/4 \\
 $m_c - m_\tau$ &  $ \cdots + c, \bar{c} $ & 289/4 \\
 $m_\tau - m_b$ &  $ \cdots + \tau, \bar{\tau} $ & 303/4 \\
 $m_b - M_Z$ &  $ \cdots + W^{+}, W^{-}, b, \bar{b}$ & 369/4 \\
  $M_Z - m_t $ &  $ \cdots + Z $ & 381/4 \\
$ > m_t $ & $ \cdots + t, \bar{t} $ & 423/4 \\
$\cdots$ & $ \cdots $   & $ \cdots $   \\
\noalign{\smallskip}\hline
\end{tabular}
%%%%$^a$ \,\, The effect of the neutrino decoupling has not been counted for. \\
%%%$^b$ \,\, The Higgs contributes +1 to $g(T)$ for $ T > m_{Higgs}$ \\
\label{tab1}
\end{table}
%%%%%%%%%%%%%%%%%%%%%%%%%%%%%%%
\\
{ {\bf{Matter Dominance :} }}
After radiation dominated era the Universe started entering the period in which matter dominated. During this period,  
the mass density was much larger than the pressure $\rho >> p$. Then in the equation for the energy conservation  the pressure term can be neglected leading to  $\;d (\rho \; a^3 ) / dt= 0\;$.  
This is solved to yield $\; \rho \simeq 1 / a^3\;$. Then from (\ref{ex3}) we have $\; a^{1/2}\; \dot a = const\;$ which is solved to yield $\;a \sim t^{2/3}$. 
During the matter dominated era the cosmic scale factor, the density and the temperature behave as 
\bea
a \sim t^{2/3} \, , \,\rho \sim t^{-2} \, , \, T \sim t^{-2/3}
\eea
%%%%%%%%
In this period the expansion rate is $H\;=\;2/3t$. 

The temperature $T_{EQ}$ at which the Universe entered the matter dominated era is 
estimated to be around $T_{EQ} \simeq \; 10^4 \; ^0K$ equivalent to $1 \; \mathrm{eV}\;$. This estimate follows by equating radiation  $\rho_{r}$ and matter density $\rho_{m}$. 
\\
{ {\bf {Vacuum Dominated Universe :}}}
Since the Universe expands and its temperature drops, eventually the cosmological term within the density in (\ref{ex3}), if it exists, will take over. Hence it is worth exploring this case too. 
If the Dark Energy of the Universe is attributed to a cosmological term with $\; \Lambda >0 \;$ this will dominate at some epoch. During this era from (\ref{ex3}), by putting 
$\rho = \Lambda$, we get 
\begin{eqnarray}
{ \left( \frac{\dot{a}}{a} \right)}^2 \,& \simeq &\, \frac{8 \pi G_N }{3} \Lambda     
\label{lam} 
\end{eqnarray}
%%%%%%%%%%%%%%
while for the acceleration it is found  
\begin{eqnarray}
  \frac{\ddot{a}}{a}  \,&=&\,  \frac{8 \pi G_N }{3}  \Lambda   \,\,. \label{lam2}
\end{eqnarray}
%%%%%%%%%%%%%%
Note that in (\ref{lam}) the r.h.s is negative for $\Lambda < 0$ while the l.h.s. is positive. We therefore conclude that in the presence of a negative cosmological constant, the Universe never enters the regime in which the cosmological term is dominant. In fact the cosmic scale factor, in this case, attains a maximum value before reaching this regime. However this is not the case when the cosmological constant is positive as we have assumed. From (\ref{lam}) we get, in this case, 
\begin{eqnarray}
a \; \simeq \; \exp \;{   \sqrt{\frac{8 \pi G_N \Lambda  }{3}} \; t} \,\,.
\end{eqnarray}
Since $\Lambda >0 $, (\ref{lam2}) implies that $\ddot a > 0$, that is the Universe is accelerated in the vacuum dominated era. In other words gravitational forces are repulsive. In this period the pressure is negative, since $p_{vac}= -\rho_{vac}= -\Lambda$. The present cosmological data show that our Universe has already entered into this phase. In fact from the values of matter and energy densities we have already seen that the deceleration parameter, see eq (\ref{decel}), is negative. 

%\subsection{\lq \lq Decoupling" or \lq \lq {Freeze-out}" of particles}
\section{Dark Matter}
In a hot Universe which is filled with particles interacting with each other, these all are in thermal equilibrium, at some epoch. However as Universe cools and expands some of these may go out of thermal equilibrium and eventually decouple. The equilibrium criterion is that the mean free path 
$l_{m.f.p.}$ is smaller than the distance these particles travel since the beginning, i.e.  
%%%
\bea
l_{m.f.p.} < v \;t \,\,. \label{mfp}
\eea
%%%%%
If (\ref{mfp}) holds, then particles will interact with the cosmic soup and there is no way of escaping. The mean free path is defined by 
$\; l_{m.f.p.} \equiv 1 / ( n \sigma ) \;$ where $\sigma $ is the interaction cross section of the particles under consideration and $n$ their density . Since the expansion rate is inversely proportional to time $H \sim 1/t$ the equilibrium criterion ($\ref{mfp}$) can be expressed as 
%%%%%%%%%
\bea
\Gamma > H \,\,\,\, , \,\,\,\, ( \; \Gamma \equiv \frac{1}{v\;n\;\sigma} \; ) \, . \label{criter}
\eea
%%%%%%%

If $\Gamma>H$ at some epoch and $\Gamma < H$ at later times, then there is a temperature $T_D$ 
for which $\Gamma = H$. $T_D$ is called the decoupling or freeze-out temperature. 
For $T \leq T_D$ these particles do not interact any longer with the cosmic soup and they expand freely. Their total number, after decoupling, remains constant and thus their density decreases with the cube of the cosmic scale factor $n \sim 1/a^3$. 

A notable example of this situation are neutrinos. They decoupled when the Universe was as hot as 
ten million Kelvin degrees and their relics today account for a small fraction of the total energy of the Universe. 
Neutrinos interact only weakly and they are nearly massless. For temperatures below the muon mass, 
that is $k T < m_\mu$, the active degrees of freedom in the hot Universe are the photons, the neutrinos and their antiparticles,  
the electrons and the positrons. The interaction cross section of neutrinos and antineutrinos with 
the electrons and positrons is  
$ \; \sigma_{\nu} \sim {\left( G_F / \sqrt{\pi} \right)}^{2} \; {( k T)}^2\;$ where $G_F$ is the Fermi coupling constant and $k T$ is the energy neutrinos carry. The total  density of neutrinos, antineutrinos, electrons and the positrons which 
interact weakly with each other is $\; n \simeq {( k T)}^{3} $, while all these are relativistic at temperatures $T >> m_e$. Therefore, their velocities are $v \simeq 1$ and 
the quantity $\Gamma $ in (\ref{criter}) is 
%%%
\bea
\Gamma \; \simeq \; {\left( G_F / \sqrt{\pi} \right)}^{2} \; {( k \;T)}^5 \,\,. \label{gam2}
\eea
At this temperature the expansion rate is 
%%%%%%%%%%
\bea
H \;=\; 1.66 \; \sqrt{g^*}\; \frac{{( k\; T)}^2}{M_{ \mathrm{Planck}}} \,\, . \label{hub2}
\eea
%%%%%%%%%%%%%%%%
where $g^*$ is the value of the function $g(T)$ at this temperature which, as read from table~\ref{tab1}, is $43/4$. The neutrino freeze-out temperature is found, by equating (\ref{gam2}) and (\ref{hub2}), to be  $ \; k \;T_D \simeq \; 2\; \mathrm{MeV}$. For $T \leq T_D$ the neutrinos and their antiparticles decouple and they do not interact 
any longer with electrons and positrons. They can be conceived, as being in an isolated  bath at temperatute $T_{\nu}$ which equals to the photons temperature at the moment of decoupling. Since they do not interact any longer with the rest of the particles, namely $\gamma, e^{-}, e^{+} $, their total number is locked. When the temperature reached $T \sim 2\; m_e$ the electrons and positrons started being annihilated to two photons through the process $ e^{-}\; e^{+} \rightarrow 2 \; \gamma $ but the photons did  not have enough energy to produce back the electrons and the positrons. Because of that, the temperature of the photons is increased ( photon reheat~) but this 
is not felt by the neutrinos since the latter do not interact with the soup. The increase of the photon temperature can be calculated from the conservation of entropy. 
%%%%%%%%%%%%%%%%%%%%   
Since the neutrinos have the same temperature with the photons  before photon  reheat we finally get 
%%%%%%%%
\bea
T_\nu \;=\; { \left( \frac{4}{11} \right) }^{1/3} \; T_\gamma \; .
\eea
%%%%%%%%%
Because of that the value of $g(T)$ in the first row of the table~\ref{tab1} should be corrected. With the effect of neutrino decoupling and photon reheat taken into account the contributions of neutrinos and photons to the energy density after photon's reheat is proportional to
$ 2 \; T_\gamma^4~+~( 21 /4 ) \; T_\nu^4 $ which gives a value equal to 
$ 2~+~(21 /4)\; {(T_\nu / T_{\gamma}) }^{4} \simeq 3.36 $ for the function $g(T)$, which is almost half of $29/4$ appearing in table~\ref{tab1}. As already remarked neutrinos are rather ruled out as Hot or Warm Dark Matter candidates and hence we will drop them from the disussion in the following. 

Various models of Particle Physics predict the existence of Weakly Interacting Massive Particles, 
called for short WIMPs, that have decoupled long ago and their densities at the present epoch may account for the \lq \lq{missing mass}" or {\bf{Dark Matter} }
of the Universe \cite{weinlee,steigman,gold,ehnos},\cite{Jungman:1995df}.
In Supersymmetric theories in which R-parity is conserved the Lightest Supersymmetric particle (LSP) is, in most of the cases, the lightest of the neutralinos $\; \tilde{\chi} \;$, a massive stable and weakly interacting particle. 
However other options are available like for instance the gravitino, the axino or the sneutrino
{\footnote{ Sneutrino is rather ruled out by accelerator and astrophysical data. For a discussion on the axino LSP cosmology and associated constraints on the reheating temperature see \cite{Brandenburg:2004du}. See also talk by G. Lazarides, Proceeding of the Third Agean School on Cosmology, Chios-Greece, 26 Sept.-1 Oct., 2005.}}
This qualifies as Dark Matter candidate provided its relics is within the experimentally determined DM relic density that is 
$\; \Omega_{\tilde \chi}\;h_0^2 \sim 0.1 \;$ ( for a review see \cite{Lahanas:2003bh} ). Its relic abundance can be calculated using the transport Boltzmann equation which will be the subject of the following section. 
 
%%%%%%%%%%%%%%%%%%%
\section{Calculating DM Relic Abundances}
{\subsection {The Boltzmann Transport Equation}}
The number density of a decoupled particle can be calculated by use of the Boltzmann transport equation. Let us assume for defineteness that the LSP particle under consideration is the neutralino, $\tilde \chi$,  although most of the discussion can be generalized to other sort of particles as well. In order to know its relic abundance and compare it with the current data we should compute  its density today assuming that at some epoch $\tilde \chi$ s were in thermal equilibrium with the cosmic soup. Their density decreases because of annihilation only since they are the LSP s and hence stable. If the $\tilde \chi$ density at a time $t$ is $n(t)$ then it satisfies the following equation known as {\emph {Boltzmann transport equation}} 
%\index{Boltzmann transport equation}
%%%%%%%%%%%%
\bea
\frac{dn}{dt}\;=\;-\;3\;\frac{\dot a}{a}\;n  \;-\; \vev{v\;\sigma} \;(\;n^2\;-\;n_{eq}^2\;) \,\,. \label{bte}
\eea
%%%%%%%%%
In (\ref{bte}), $\sigma$ is the cross section of the annihilated $\tilde \chi$ s, and $v$ is their relative velocity. The thermal average $\vev{v \sigma}$ is defined in the usual manner as any other thermodynamic quantity. 

The first term on the r.h.s.  of (\ref{bte}) is  easy to understand. It expresses the fact that the density changes because of the expansion. If we momentarily ignore the interactions of the $\tilde \chi$ s with the rest of the particles then their total number remains constant. Therefore, $n\;a^3 = \mathrm{const.}$ from which it follows,  by taking the derivative with respect the time, that the density rate is given by the first term on the r.h.s. 
of (\ref{bte}). However the $\tilde \chi$ s do interact and their number decreases because of pair annihilation. In fact LSP's are stable Majorana particles and it is possible to be annihilated by pairs to Standard Model particles.
Their number is therefore reduced until the freezing-out temperature below which they do not interact any further with the remaining particles and their total number is locked. 
%%%%%%%%
{\footnote{ If other unstable supersymmetric particles are almost degenerate in mass with the 
$\tilde \chi \;$s they are in thermal equilibrium with these almost until the decoupling temperature affecting the relic abundance of $\tilde \chi \;$s through the mechanism of the 
{\emph {co-annihilation}}. We drop momentarily this very interesting case from the discussion. }} 
%%%%%%%%%%%
Therefore, their density decreases as $dn / dt \;=\; - \;n \Gamma_{ann}$, where the annihilation rate $\Gamma_{ann}$ is given by $\Gamma_{ann} \;=\; v\;\sigma\;n\;$. This explains the second term on the r.h.s. of (\ref{bte}). However the $\tilde \chi$ s do not only annihilate but are also produced through the inverse process. 
The last term on the r.h.s of (\ref{bte}) expresses exactly this fact. Note that 
when $\tilde \chi$ s were in thermal equilibrium with the rest of the particles and the environment was hot enough, the  annihilated products had enough energy to produce back the $\tilde \chi$ s at equal rates. During this period $n\;=\;n_{eq}$ and the last two terms on the r.h.s. of (\ref{bte}) cancel each other, as they should. 

Thus the picture is the following.
The $\tilde \chi$ s are in thermal and chemical equilibrium at early times. During this period $\Gamma >> H$, 
 see (\ref{criter}), and $n = n_{eq}$. However as the temperature drops 
and eventually passes  $k \;T \sim \; m_{\tilde \chi}$ the $\tilde \chi$ s annihilate but their products do not have enough thermal energy to produce back the annihilated $\tilde \chi$ s. The $\tilde \chi$ s are in thermal but not in chemical equilibrium any 
more. In addition their density drops exponentially $\exp \;( - m_{\tilde \chi}/k T)$ following the Boltzmann distribution law and $\Gamma \equiv n v \sigma$ decreases so that eventually at a temperature $T_f$, the freeze - out temperature, $\Gamma$ equals to the expansion rate $H$. Below this temperature $\Gamma < H$ and the $\tilde \chi$ s are out of thermal equilibrium. They decouple not interacting any longer with the cosmic soup and they expand freely. 
Their total number is locked to a constant value and their density changes because of the expansion. Actually for $T < T_{\tilde \chi}$ their density is much larger than the equilibrium density $n >> n_{eq}$ and since $\Gamma >> H$, the first term dominates in (\ref{bte}) so that $n\;a^3$ is indeed a constant.
%%%%%%%%%%%%%%%

Concerning the cross section thermal average,  
in general for two annihilating particles $1,2$, under the assumption that they obey  Boltzmann statistics, which is valid for $T \leq m_{1,2}$,  
%%%%%%%%%%%%%
{\footnote{
The initial condition in solving Boltzmmann's equation should lie in this regime and at a point above the decoupling temperature. This is perfectly legitimate for the case of neutralinos since their decoupling temperature is well below $m_{\tilde \chi}$, $T_f \sim m_{\tilde \chi} /20\; $ .
}}
%%%
one finds \cite{gondolo,gondolo2} 
%%%%%%%%%%%%%%%%%%%%%%%%%%%
\beq
\vev{v \sigma_{12}}\;=\; 
\frac{ \int_{{(m_1+m_2)}^2}^{\infty}\; ds\;K_1(\sqrt{s}/T) \; p_{cm}\;W(s)}
{2 \;m_1^2\; m_2^2 \;T\;K_2(m_1/T)\;K_2(m_2/T) } \; \label{saverage}
\eeq
%%%%%%%%%%%%%%%%%%%%%%%%%%%
where $p_{cm}$ is the magnitude of the momentum of each incoming particle in their CM frame and $K_{1,2}$ are Bessel functions. The quantity $W$ within the integral is related to the total cross section $\sigma (s)$ through 
$\; \sigma=4 \; p_{cm} \; \sqrt{s} \;W(s) /\lambda (s,m_1^2,m_2^2)\;$  
%%%%%%%%%%%
{\footnote{
$ \lambda(x,y,z)\; \equiv \; x^2 + y^2 + z^2 - 2\;(x y + y z + z x ) \;. $
}}.
%%%%%%%%%%%%%
This expression can be considerably simplified if one follows a non-relativistic treatment expanding the cross section in powers of their relative velocity $\;v\;$,
%%%%%%%%%%%
\beq
v \; \sigma \;=\;a \;+\; \frac{b}{6} \;v^2 \; \;.
\eeq
%%%%%%
With this approximation the LSP s annihilation thermal average is
%%%%%%%%%%%
\beq
\vev {v \; \sigma } \;=\;a \;+\; ( b - \frac{3}{2}\; a )\; \frac{k T}{m_{\tilde \chi}}\; .
\eeq
%%%%%%
The non-relativisatic expansion is legitime in energy regions away from poles, some of which may be of particular physical interest, and thresholds as well. However near such points this approximation behaves badly invalidating physical results. Thus we had better used the thermal average in the form given by the Eq. (\ref{saverage}).

The goal is to solve (\ref{bte}), in order to know the density at today's temperature $T_0 \simeq 2.7 \; \;^0K$, provided  that $n=n_{eq}$ long before decoupling time. By defining 
$\; Y \equiv n/s \;$, where $\;s\;$ is the entropy density,  and using a new variable $\; x= T/m_{\tilde \chi} \;$
%%%%
{\footnote{We use now units in which the Boltzmann constant $k$ is unity, or same we absorb it within the temperature $T\;$ . }}
%%%%
, one arrives at a simpler looking equation 
%%%%%%%%%%%
\bea
\frac{dY}{dx}\;=\;
m_{\tilde \chi} \vev{v \;\sigma} {\left( \frac{45 G_N}{\pi} \; g \right)}^{-1/2}
(\;h + \frac{x}{3 }\; \frac{dh}{dx}  \;) \;(\;Y^2\;-\;Y_{eq}^2\;) \,\, . \label{bte2}
\eea
%%%%%%%%
The function $h$ that appears in this equation counts the effective entropy degrees of freedom related to the entropy density through $s= k \; \frac{2 \pi^2}{45}\; {(k T)}^3 h(T)$.
The prefactor of $\;Y^2\;-\;Y_{eq}^2\;$ is usually a large number, due to the appearance of the gravitational constant $G_N^{-1/2}$, and this can be exploited in using numerical approximations reminiscent of the WKB in Quantum Mechanics \cite{Lopez:1991pp,Lahanas:1999uy}. \\
One can solve this to find $\; Y\;$ today $\;T_{\gamma} \;$ which is very close to $\;Y_0=Y(0)\;$,  and from this the matter density of $\;\tilde \chi \;$s. The latter is 
%%%%%%%%%%%%%%%
\beq
\varrho_{\tilde \chi}\;=\;n_{\tilde \chi} \; m_{\tilde \chi}\;=\;m_{\tilde \chi}\; s_0 Y_0 \;=\;
\frac{2 \pi^2}{45} \;m_{\tilde \chi} \;h_{eff}^0 \;Y_0\; T_{\gamma}^3 \; , \label{rhox}
\eeq
%%%%%%%%%%%%
where $h_{eff}^0 \equiv h(T_{\gamma})$ is today's value for the effective entropy degrees of freedom which is 
$\; h_{eff}^0   \simeq 3.918 \;$
%%%%%%%%%%%%%%% 
{\footnote{ 
At today's temperatures only photons and neutrinos contribute to $h_{eff}$. Its value is almost half of $\; 2+6 \times \frac{7}{8}= 7.25 \;$ one naively expects by merely counting the spin degrees of freedom of the photons and neutrinos due mainly to the decoupling of the neutrinos. 
}}.
%%%%%%%%%%%%%%%%%%%%%%%%
The relic density is then
%%%%%%%%%%%%%%%%%%%%%%%%%%%
\bea
\Omega_{\tilde \chi} \; {h_0}^2 \;= \; {h_0}^2 \; \frac{\varrho_{\tilde \chi}}{\varrho_c} 
\;=\;0.6827 \times 10^8 \; \frac{m_{\tilde \chi}}{GeV} \;h_{eff}^0 Y_0 \; 
{\left( \frac{T_{\gamma}}{T_0}\right)}^3 \; . \label{omegax}
\eea
%%%%%%%%%%%%%%%%%%%%%%%%%%
In writing (\ref{omegax}) we use the fact that $\varrho_c=8.1 \times 10 ^{-47}\; h_0^2\; GeV^{4}$. We have also expressed the dependence on temperature through the ratio $T_{\gamma}/{T_0}$ by using a reference temperature $T_0=2.7 \; ^0 K$. The CMB temperatute has been determined by measurements of the CMB to be $T_{\gamma}=2.752 \pm 0.001 \; ^0 K $. The value of $Y_0$ required can be found by solving 
Eq. (\ref{bte2}) numerically with the boundary condition that $Y \rightarrow Y_{eq}$ at temperarures well above the freeze-out temperature. The numerical solution proves to be rather time consuming and for this reason many authors use approximate solutions which are less accurate, by only 5-10 \%, having the advantage that the calculation performs fast and the physical content is more transparent. There are good packages in the literature, like DarkSUSY \cite{darksusy} and microOMEGAs \cite{momegas}, which can be used to handle numerically the Boltzmann equation and find the LSP relic density in supersymmetric theories.

%%%%%%%%%%%%%%%%%%%
{\subsection{ Appxoximate Solutions to Boltzmann equation}}
Approximate solutions can be found under the assumption that $\; Y \ll Y_{eq} \; $ below the freezing point $\;x_f = T_f / m_{\tilde \chi} \;$ while $\; Y \simeq Y_{eq} \; $ above it. Omitting then the $Y_{eq}$ term in  (\ref{bte2}), which is valid for $x$ between $x_f$ and $x_0 \simeq 0 $, and putting $x=x_f$ in (\ref{bte2}) an equation for $x_f$ is derived 
%%%%%%%%%%%%%%%%%%%%%%%%%%%%
\beq
x_f^{-1}\;=\;\ln \;[ \;0.03824 \; g_s\; \frac{M_{Planck} \;m_{\tilde \chi}}{\sqrt{g^{*}}} \;
\vev{v\;\sigma} \; c(c+2) \; x_f^{1/2} \;]  \; , \label{freezep}
\eeq
%%%%%%%%%%%%%%%%%%%%%%%%%%%
which can be solved numerically to obtain $x_f$. In this $g^{*}$ stands for the effective energy degrees of freedom at the freeze-out temperature $\;g^{*}= g(x_f)\;$ and the derivative term $\; dh/dx \;$ in (\ref{bte2}) has been ignored in this approximation. $g_s$ are the spin degrees of freedom and the $c(c+2)$ within this expression equals to one. The reason we present it in the above equation is that empirically it is found that very good approximation for the freeze-out point temperature is obtained with values $\; c \simeq 1/2 \;$. The freeze out temperature for a WIMP is close to $T_f \simeq m_{\tilde \chi}/20 \;$. Now that $x_f$ has been determined one can integrate 
(\ref{bte2}) from $x_f$ to zero, under the same assumptions, to obtain $Y(0)$ and from this the relic abundance. The solution for $Y(0)$ entails to a density
%%%%%%%%%%%%%%%%%%%%%%%%%
\beq
\varrho_{\tilde \chi}\;=\;{(\;\frac{4 \pi^3}{45}\;)}^{1/2} \; 
{ \left( \;\frac{T_{\tilde \chi}}{T_{\gamma}}\;\right)}^3 \; \frac{T_{\gamma}^3}{M_{Planck}} \;
\frac{\sqrt{g^{*}}}{J}  \; . \label{rrrx}
\eeq
%%%%%%%%%%%%%%%%%%%%%%%%%%
In (\ref{rrrx}) the quantity $J$ is given by the integral $\; J \equiv \int_0^{x_f} \vev{v\sigma} \;dx \;$. Recall that $g^{*}= g(x_f)$ in the notation we follow here. In this expression the $\tilde \chi$'s temperature $T_{\tilde \chi}$ appears explicitly which is different from that of photons. In fact it is found \cite{schramm} that  due to the decoupling of both $\; \tilde \chi \;$s and neutrinos that 
$\; {({T_{\tilde \chi}}/{T_{\gamma}})}^3 = {4}/{11} \cdot {g(T_\nu)}/{g(T_f)} =
{3.91}/{g(T_f)}$ with $T_\nu$ the neutrino decoupling temperature and $T_f$ that of 
$\; \tilde \chi \;$s. Using this and (\ref{rrrx}) we obtain for the relic density 
$\Omega_{\tilde \chi} \; {h_0}^2 =  {\varrho_{\tilde \chi}}/{( 8.1 \times 10^{-47} \; GeV^4 )} \;$ the 
following result
%%%%%%%%%%%%%%%%%%%%%%
\beq
\Omega_{\tilde \chi} \; {h_0}^2 \;=\; \frac{1.066 \times 10^9 \; GeV^{-1}}{M_{Planck} \; \sqrt{g^*} \;J}
\; , \label{omegaap}
\eeq
%%%%%%%%%%%%%%%%%%%%%
which is the expression quoted in many articles. The quantity $J$, which has already been defined,  is  given in $GeV^{-2}$ units. Note that the relic density is roughly inverse proportional to the total cross section. This means that the larger the cross section the smaller the relic density is and vice versa. 
 
 To have an estimate of the predicted  relic density for the case of a neutralino LSP we further approximate $J \approx x_f \; {\vev{v\;\sigma}}_f$ and $\; \vev{v \sigma} \sim \alpha / m_{\tilde \chi}^2 \;$ where $\alpha$ is a typical electroweak coupling. Since $x_f$ is of the order of 
$ \sim 0.1 $ and $g^*$ is $\; \simeq 100\;$ for masses in the range $m_{\tilde \chi} \simeq 20\;GeV - 1 \;TeV$, the relic density (\ref{omegaap}) above turns out to be $\Omega_{\tilde \chi} \;h^2_0 \sim 0.1$ in the physically interesting region $m_{\tilde \chi} \simeq \; 100 \;GeV$.  Therefore we 
conclude that relic densities of the right order of magnitude can naturally arise in supersymmetric theories if one interprets the Dark Matter as due to a stable neutralino. \\

%%%%%%%%%%%%%%%%%%%%%%%%%%%%%%%%%%%%%%%
{\subsection{Co-annihilations}}

All disussion so far concerned cases in which the stable  WIMP is not degenerate in mass with other heavier species that can decay to it. Therefore there is an epoch where the Universe is filled by SM particles and the LSPs, whose density decreases because of pair annihilations when  the temperature starts passing the point where SM particles do not have enough energies to produce back LSPs. However it may happen that although lighter the LSP's mass $m_{\tilde \chi}$ is not very different from other particles's masses $m_i$ that they decay to it. In fact when $\delta m_i \equiv m_i - m_{\tilde \chi} \sim T_f$ these particles are thermally accessible and this implies that they are as  abundant as the relic species. This drastically affects the calculation of the LSP relic density. This effect is known as co-annihilation 
\cite{coannih1,griest,gondolo2,edsjo,coannih2} and it is not of academic interest. Actually in the most popular supersymmetric schemes, advocating the existence of good  CDM candidates, there are regions having this characteristic so it is worth discussing this case. 

Since all nearly degenerate particles with the LSP will eventually decay to it the relevant quantity to calculate for the relic abundance of the LSP is the  density $n = \sum_i \; n_i \;$. In it $n_i$ is the density of the particle $i$ and the sum runs from $i=1,...N$. With $i=1$ we label the LSP and with $i=2,...N$ the rest of the particles that are almost degenerate in mass with it. Following \cite{griest} the Boltzmann transport equation (\ref{bte}) is generalized to,   
%%%%%%%%%%%%
\bea
\frac{dn}{dt}\;=\;
-\;3\;\frac{\dot a}{a}\;n  \;-\; \sum_{i,j}^N \vev{v_{ij}\;\sigma_{ij}} \;
(\;n_i n_j \;-\;n^{eq}_i n^{eq}_j \;) \,\,. \label{btecoan}
\eea
%%%%%%%%%
The notation in Eq. (\ref{btecoan}) is obvious.
Since the criterion for which particles co-annihilate with the LSP is roughly given by 
$\delta m  \sim T_f$ and $T_f \simeq m_{\tilde \chi}/20 $ the particles participating in the co-annihilation process are those for which the  mass differences are  
$\delta m_i \simeq 5 \% \; m_{\tilde \chi} $. Their equilibrium densities are given by 
$\; n_i^{eq}=g_i \; \frac{T}{2 \pi} \; m_i^2 K_2(m_i/T) \;$ where $g_i$ denotes the spin degrees of freedom. Approximating $\; n_i/n \simeq n_i^{eq}/n^{eq} \; $,  Eq. (\ref{btecoan}) can be cast in the following form
%%%%%%%%%%%%
\bea
\frac{dn}{dt}\;=\;
-\;3\;\frac{\dot a}{a}\;n  \;-\; \sum_{i,j}^N \vev{v\;\sigma_{eff}} \;
(\;n^2 \;-\;n_{eq}^2 \;) \,\,. \label{btecoan2}
\eea
%%%%%%%%%
where the effective thermal average appearing in this equation is a generalization of (\ref{saverage}) given by
%%%%%%%%%%%%%%%%%%%%%%%%%%%
\beq
\vev{v \sigma_{eff}}\;=\; \sum_i^N \frac{n_i^{eq} \; n_j^{eq}}{n_{eq}^2} \; \vev{v_{ij} \sigma_{ij} }\;.
\label{saverage2}
\eeq
%%%%%%%%%%%%%%%%%%%%%%%%%%%
This can be written as a single integral generalizing the results of reference \cite{gondolo}
%%%%%%%%%%%%%%%%%%%%%%%%%%%
\beq
\vev{v \sigma_{eff}}\;=\; 
\frac{ \int_{2}^{\infty}\; da\; K_1(a/x) \sum_{i,j} \lambda(a^2,b_i^2,b_j^2) \;g_i g_j\; \sigma_{ij}(a) }  
{4 x \;{( \;\sum_i  g_i b_i^2 K_2(b_i/x) \; )}^2} \; , \label{saverage3}
\eeq
%%%%%%%%%%%%%%%%%%%%%%%%%%%
where $\; b_i \equiv m_i / m_{\tilde \chi} \;$ with $\tilde \chi$ denoting the LSP labellel by 
$i=1$. If we seek for an approximate solution, as was done in the no co-annihilation case, then the freeze-out point is 
%%%%%%%%%%%%%%%%%%%%%%%%%%%%
\beq
x_f^{-1}\;=\;\ln \;[ \;0.03824 \; g_{eff}\; \frac{M_{Planck} \;m_{\tilde \chi}}{\sqrt{g^{*}}} \;
\vev{v\;\sigma_{eff}} \; c(c+2) \; x_f^{1/2} \;]  \; . \label{freezepco}
\eeq
%%%%%%%%%%%%%%%%%%%%%%%%%%%
In this $g_{eff}$ is defined as 
%%%%%%%%%%%%%%%%%%%
\beq
g_{eff}\;\equiv \; \sum_i \; g_i \;{(\; 1 + \Delta_i \;)}^{3/2} \; \exp{( - \Delta_i /x_f )}
\eeq
%%%%%%%%%%%%%%%%%%%
where $\; \Delta_i \equiv (m_i-m_{\tilde \chi}) / m_{\tilde \chi}\;$. As for the relic abundance, this is given by (\ref{omegaap}) with the quantity $J$ defined now as 
$\; J \equiv \int_0^{x_f} {\vev { v \sigma_{eff} } } \;dx  \;$. In calculating this we have made use of the fact that all nearly degenerate particles with the LSP will eventually decay and therefore the final LSP abundance will be extracted from $\; n = \sum_i \; n_i \;$. 
In supersymmetric models the relevant co-annihilation channels are between neutralinos, charginos and sfermions. 

As a preview of the importance of the co-annihilation process we mention that the cosmological bound on the LSP neutralino is pushed to $\; \simeq 600 \;GeV\;$, from about $\;200\;GeV$ in the stau co-annihilation region and to $\; 1.5 \;TeV$ in the chargino co-annihilation, increasing upper bounds and weakening the prospects of discovering supersymmetry in high energy accelerators.
\\ 
%%%%%%%%%%%%%%%%%%%%%%%%%%%%%%%%%%%%%%%%%%%%%%%
%%%%%%%%%%%%%%%%%%%SUPERSYMMETRY

\section{Supersymmetry and its Cosmological Implications}

Supersymmetry, or SUSY for short, is a fermion-boson symmetry and it is an indispensable ingredient of Superstring Theories. Fermions and bosons go in pairs ( partners ) having similar couplings and same mass~!. Spontaneous Symmetry Breaking ( SSB ) of local SUSY at Planckian energies makes the partner  of the graviton, named gravitino, massive 
$\; m_{3/2}\; \neq 0 \;$ ( {\bf{ SuperHiggs effect}}) and lifts the mass degeneracy between  fermions and bosons by amounts $\; M_{S} \;$, a parameter which depends on the particular supersymmetry breaking mechanism. At much lower energy scales $\;E \ll M_{Planck} \;$,  accessible to LHC if $\;M_S \simeq {\cal O}( TeV ) \;$, the theory is supersymmetric but there appear terms that break SUSY softly. These are scalar mass terms, gaugino mass terms or scalar trilinear couplings 
$\; m_0, \;M_{1/2},\; A_0 \;$ ( for a review  see \cite{hnl}).  
Supersymmetric extensions of the SM naturally predict the existence of Dark Matter candidates which is the LSP. This may be the \lq \lq {Neutralino}" or the ''Gravitino", whichever is the lightest, or other non-SM particle provided its relic abundance is within the cosmological 
limits while all accelerator bounds are respected.  

In supersymmetric theories the generators of the fermion-boson symmetry are spinorial 
operators $\; Q, \; Q^{\dagger}\;$ which turn a boson state to a fermion and vice versa, 
%%%%%%%%%
\bea
Q^{\dagger} \; \ket{\, \, \, boson \, \, \; \,} \;&=&\; \ket{fermion} \\
Q \; \; \ket{fermion} \;&=&\; \ket{\, \, \, boson \, \, \; \,} \,\,.
\eea
%%%% 
These commute with the supersymmetric Hamiltonian $H_{S}$, 
%%%%%%%%%%%
\bea
[ H_S, Q] \;=\; [ H_S, Q^{\dagger}] \;=\;=0
\eea
%%%%   
resulting to a degenerate mass spectrum between fermions and bosons. The field content of a supersymmetry theory is larger than in ordinary theories in the sense that additional degrees of freedom are needed, the so called \lq \lq {sparticles}",  that are superpartners of the known particles. 
%%%%%%%%%%%%%%%%%%%%%%%%%%%%%%%%%
\begin{table}
\centering
\caption{Particle content of the Minimal Supersymmetric Standard Model}
\begin{tabular}{clccclc}
\hline\noalign{\smallskip}
& SM particles && & & SUSY particles  & \\ 
\noalign{\smallskip}\hline \noalign{\smallskip}
 Particle & \,  Name & Spin &   & Sparticle & \, \, Name  & Spin    \\  
\noalign{\smallskip}\hline \hline \noalign{\smallskip}
$q$   & \, quarks   & 1/2  &    &  $\tilde{q} $    & \, \, squarks & 0  \\
$l$   & \, leptons  & 1/2  & \, \quad $Q,\, Q^{\dagger}$ \quad \,  &  $\tilde{l} $    & \, \, sleptons   & 0  \\
$W^{\pm}, \; W^0$ & \,W-bosons & 1    &\, \quad $\Longleftrightarrow$  \quad \, &  ${\tilde{W}}^{\pm}, \; {\tilde{W}}^0 $ &  \, \, winos  & 1/2 \\
$B$   & \, B-boson       &  1  &   & $\tilde{B} $ &\, \,  bino  & 1/2 \\
$G$   &\, gluons        & 1   &     &  $\tilde{G} $        & \, \, gluinos     & 1/2 \\
$ H_{1,2}$  & \,Higgses   & 0   &     &  $ {\tilde{H}}_{1,2}$   &\, \,  Higgsinos       & 1/2 \\
\noalign{\smallskip}\hline 
\end{tabular}
\label{tab2}
\end{table}
%%%%%%%%%

The most economic supersymmetric extension of the Standard Model, known as MSSM ( Minimal Supersymmetric Standard Model ) has the physical content appearing in  table (\ref{tab2}). 
In it the SM particles are shown on the left with their corresponding 
spins and their superpartners on the right. Note that unlike the SM the Higgses $H_{1,2}$ are 
not independent and thus  five Higgses survive the Electroweak Symmetry Breaking.  
In addition MSSM can posses a symmetry known as R-parity under which each particle 
bears a quantum number 
%%%%%%%%%
\bea
P_R \;=\; {(\; -1)}^{3\;(B-L)\;+\;2\;s} \,\,,
\eea 
%%%%%%%%%
with $B,L$ the baryon and lepton number of the particle and $s$ its spin. $P_R$ is $\;+1$ for particles and $\;-1$ for the superparticles. This quantum number is multiplicatively conserved in theories possessing R-parity and prohibits Baryon and Lepton number violations. Another virtue of this symmetry is the fact that the lightest supersymmetric particle (LSP) is stable. The reason for this is that, in R - parity conserving theories, the vertices have an even 
number of sparticles. Because of this, a sparticle can only decay to an odd number of sparticles and an even or odd number of SM particles. For the LSP such a decay is however energetically forbidden since it is the lightest sparticle and hence it is stable.
If, in addition, it is electrically neutral and does not interact strongly the LSP qualifies as a WIMP.

The CMSSM ( constrained MSSM ) is the most popular and extensively studied supersymmetric model encompassing SM. It is motivated by the minimal supergavity theories ( mSUGRA ) and differs from MSSM in that universal boundary conditions are imposed for the scalar, gaugino and trilinear couplings at a unification scale so that there is a single $m_0$, $M_{1/2}$ and $A_0$. These along with the ratio $\vev{H_2}/\vev{H_1}$ can be chosen to be the only arbitrary parameters of the theory. Other parameters $\mu, m_3^2$, inducing mixing between the Higgs multiplets are determined from the electrowek symmetry breaking conditions
{\footnote{
The sign of  $\mu$  is also a free parameter. 
}}.
%%%%%%%%%%%%%%%%%%%%%%%%%%

\subsection{  The Neutralino DM}
The neutral components of the Higgsinos  ${\tilde{H}}_{1,2}^0$, 
the neutral \lq \lq wino" $\tilde{W}^0$  and the \lq \lq{Bino}" $\tilde{B}$ 
( see table \ref{tab2} ) interact weakly but they are not mass eigenstates. The four mass eigenstates $\; {\tilde \chi}^0_i \;i=1,...4 \;$, named \lq \lq{Neutralinos}", are linear combinations of these. 
These are Majorana fermions which means that particles are same as their antiparticles and thus they possess half the degrees of freedom of a ''charged" Dirac particle like the electron for instance. 

In the MSSM, briefly discussed in the previous section, and depending on the inputs for the SUSY breaking parameters $m_0,\;M_{1/2},\;A_0$ and $tan \beta$, the lightest of the neutralinos, which we shall denote by $\tilde {\chi}$, may be the LSP. Being a linear combination of the Higgsinos, Wino and Bino fields this can be written as 
%%%%%%%%%%%%%%
\bea
\tilde {\chi} \;=\; a_1\; \tilde{H}_1^0 \;+\;a_2 \; \tilde{H}_2^0  \;+\;a_W \; \tilde{W}^0
\;+\;a_B\;\tilde{B}  \,\,. \label{lsp}
\eea
%%%%%%%%%
where normalization requires that $\sum_i \;{| a_i |}^2 \;=\;1$. 
Depending on the magnitudes of $a_i$ appearing in (\ref{lsp}) we can distinguish the following two cases 
%%%%%%%%%%%
%\bea
\begin{eqnarray}
{| a_1 |}^2\;+\;{| a_2 |}^2\; & >> &\;{| a_W |}^2\;+\;{| a_B |}^2 \;, 
\; \; \; \; (\;\mathrm{Higgsino-like} \;)  \nonumber \\
{| a_1 |}^2\;+\;{| a_2 |}^2\; & << &\;{| a_W |}^2\;+\;{| a_B |}^2 \; , \,\, 
\; \; \; \; (\;\mathrm{gaugino-like} \;) \; \; . %\nonumber
\end{eqnarray}
%\eea
%%%%%%%%%%%%%%%%%%%%%%
In the upper case the LSP is mostly Higgsino  and in the lower mostly Gaugino. Other cases encountered in model studies are somewhere in between.  
In the CMSSM, the $\tilde \chi$ is gaugino - like, actually a bino, in the major portion of the parameter space 
{\footnote{
There are narrow regions in the parameter space, like for instance the so-called Hyperbolic Branch,  which are of particular phenomenological interest and in which the 
LSP may be a Higgsino.  
}}.  
Absence of its detection in  accelerator experiments puts a lower bound on its mass, $m_{\tilde \chi} > 46 \; \mathrm{GeV}$. 

If $\tilde \chi$ is the LSP, and thus stable in R - parity conserving theories, then at some epoch, the cosmic soup contains the $\tilde \chi$'s and Standard Model particles. All other supersymmetric particles have already  decayed to $\tilde \chi$ and SM particles. Then the number of the LSP particles can only decrease through pair annihillations to SM particles through 
the reactions 
$\tilde \chi \;+\; \tilde \chi \; \rightarrow \; A \;+\; B \; +\; \cdots $. In the MSSM and in leading order in the coupling constants involved, only two body pair annihilations take place and  the SM particles in the final state occur in the combinations displayed in table 
\ref{tab3}. As already remarked, in the minimal supersymmetric extension of the Standard Model, there exist five Higgs mass eigenstates $H^{\pm}, H, h, A$, the last three being neutral, in contrast to the Standard Model where only one neutral Higgs survives after Electroweak Symmetry breaking. We should mention that an upper theoretical bound on the lightest of the neutral Higgses mass $m_h$ exists which is $\approx 138\; \mathrm{GeV}$. 
%%%%%%%%%%%%%%%%%%%%%%%%%%%%%%%%%
\begin{table}
\centering
\caption{ Neutralino pair annihilations {$\lsp \lsp \rightarrow A+B$}}
\begin{tabular}{ll}
\hline \noalign{\smallskip}
\quad \quad \quad  Particles in the final state $\mathrm{ \;A\;,\; B\;}$    \\
\hline \noalign{\smallskip}
${\mathrm{Fermion - Antifermion \; :}}$      & $q \; \bar{q} $    \\
                                & $ l \; \bar{l} $   \\
\hline\hline\noalign{\smallskip}
${\mathrm{  \; Gauge \; Bosons \;: }}$  & $W^+\; W^- $ \\
                      & $Z\;Z $  \\   
\hline\noalign{\smallskip}  
${\mathrm{ \; Gauge \; Bososn +1 Higgs \;:}}$ & $ W^{\pm}\;H^{\mp}$ \\
                                & $Z\;A $   \\
                                & $ Z \; H $    \\   
                                & $ Z \; h$ \\   
\hline\noalign{\smallskip}
${\mathrm{ \; Higgses\; :}}$   & $ H^{+}\;H^{-} $ \\
               & $ H\;H  $\\
              &  $h\;h $ \\   
              &  $H\;h $ \\   
              &  $A\;H \;,\; A\;h $ \\   
              &  $A\;A $ \\   
\hline
\end{tabular}
\label{tab3}
\end{table}

%%%%%%%%%%%%%%%%
Accelerator experiments impose various constraints on sparticle masses which along with the cosmological bound on the DM relic abundance restrics the allowed parameter space describing the model. The potential of discovering supersymmetric particles in future experiments depends on the bounds put on sparticle masses and these are constrained by the cosmological data. LEP and Tevatron colliders give a lower bound of $104 \; GeV$ for  the chargino mass while sleptons should not weigh less than about $99 \; GeV$. 
The bounds imposed on all sparticles can be traced in \cite{LEP2}. LEP has also provided us with the important Higgs mass bound $m_h > 114 \; GeV$ \cite{higgsLEP}. Other important constraints stem from the decay $\; b \rightarrow s \; \gamma$ whose branching ratio should lie in the range 
$\; 1.8 \times 10^{-4} < BR(b \rightarrow s \; \gamma) < 4.5 \times 10^{-4} \;$ 
at the $2\;\sigma$ level \cite{bsg1,bsg2}.  The BNL E821 experiment derived a very precise value for  the anomalous magnetic moment of the muon $\;\alpha_\mu \equiv ( g_\mu-2)/2 =11659203(8) 
\times 10^{-10} \;$ \cite{E821ex} pointing to a discrepancy between the SM theoretical prediction and the experimental value given by $\; \delta \alpha_\mu =(\;361 \pm 106\; ) \times 10^{-11} \;$ which shows  a $3.3 \sigma$ deviation. This can put severe upper bounds on sparticle masses. 
However theoretical uncertainties due to disagreement between the $e^+ e^-$ and $\tau$ decay data used to calculate the contributions to $\;g_\mu-2\;$ forces us to consider these data with a grain of salt until the disagreement between the two theoretical approaches is finally resolved \cite{E821th}. 

\begin{figure}[t]
\centering
\includegraphics[width=5.7cm]{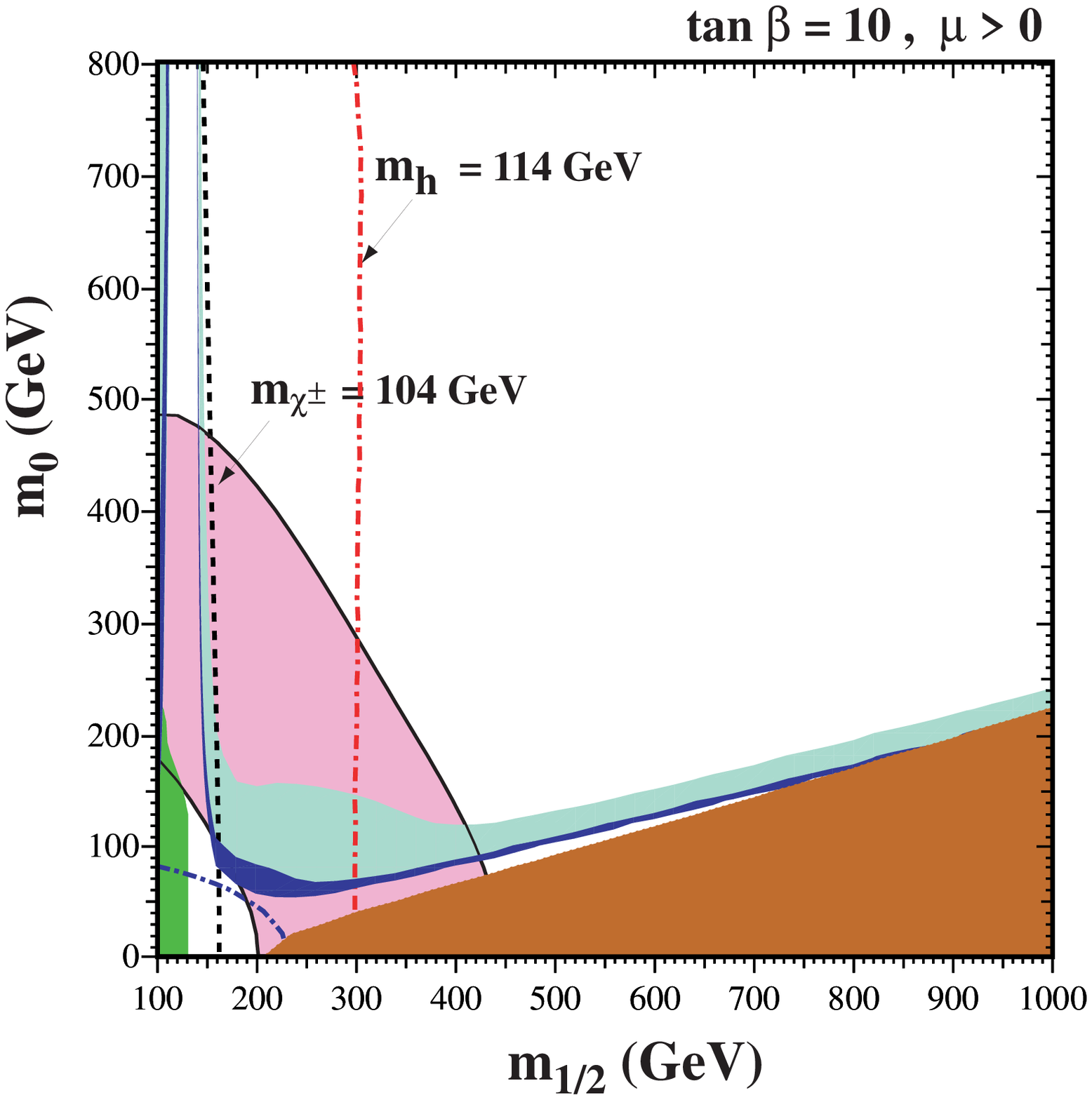} 
\hspace*{.1cm}
\includegraphics[width=5.7cm]{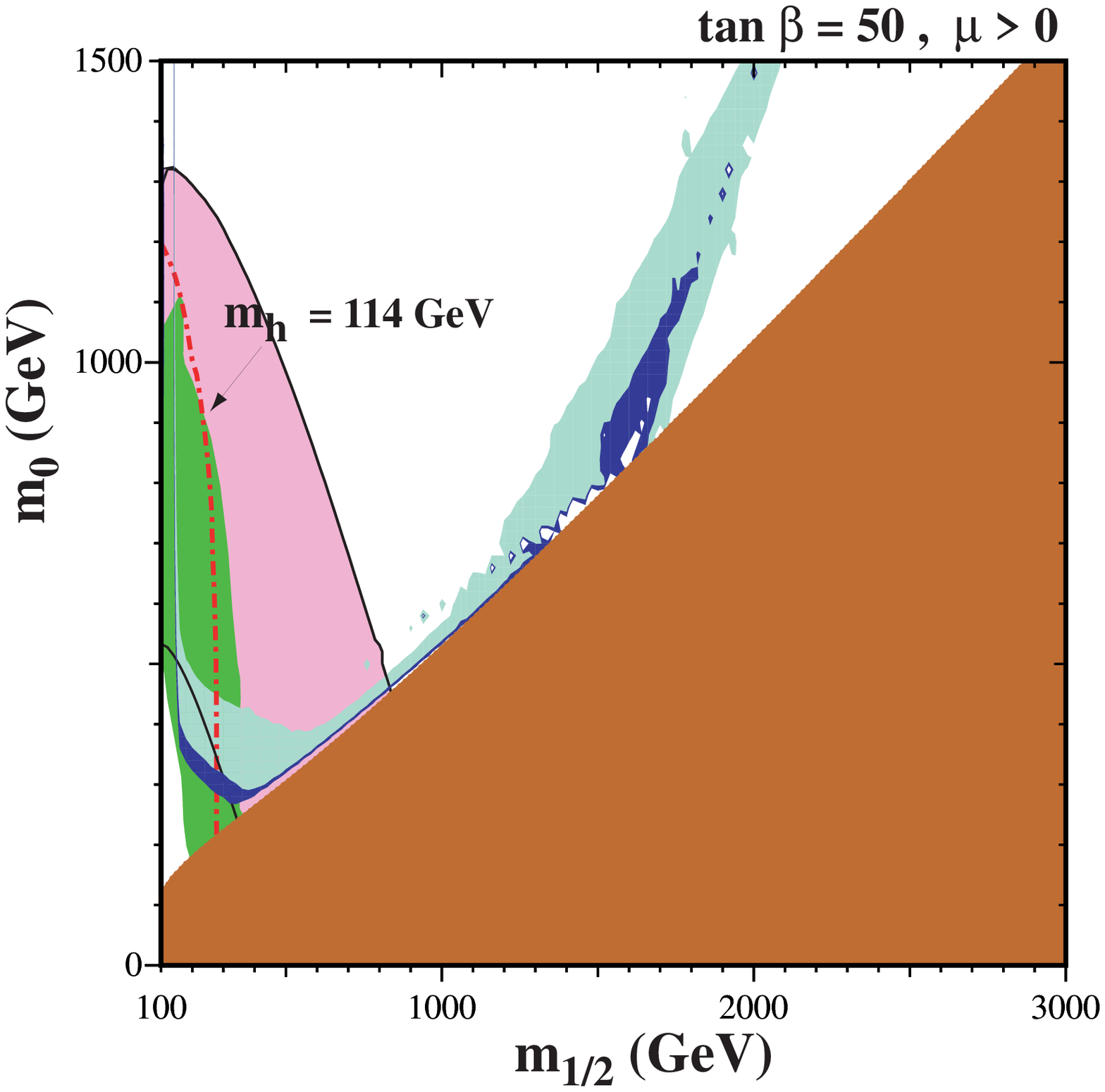} 
\caption {
mSUGRA/CMSSM constraints after WMAP.
The very dark shaded region (dark blue) is favoured by WMAP
( $0.094 \le \Omega_\chi h^2 \le 0.129$ ). 
In the medium shaded region (turqoise) $0.1 \le \Omega_\chi h^2 \le 0.3$.
The shaded region at the bottom ( brick red ) is excluded because LSP is charged.
Dark regions (green) on the left are excluded by $b \to s\gamma $.
The shaded stripes ( pink ) on the left  are favoured by $g_\mu - 2$ at the $2-\sigma$ 
level. The LEP bounds on the chargino mass $104 \;GeV$ and the Higgs mass $114\;GeV$ are also shown. ( from \cite{EllisWmap}. }
\label{eoss}
\end{figure}

There are numerous phenomenological analyses by various groups and regions in the parameter space which are compatible with cosmological and accelerator data have been delineated. 
Three main regions have been identified as conforming with all data. The ``funel" region 
$m_0 \sim M_{1/2}$ in which neutralinos rapidly annihilate via direct s-channel pseudoscalar Higgs poles, which opens up for large values of $\;tan \beta \;$, 
$\; {\tilde \chi}, {\tilde \tau} \;( \mathrm{stau} )\;$ co-annihilation region which extends to 
large $M_{1/2} >> m_0$ and the hyperbolic branch (HB), which includes the ``focus point" region, in which  $m_0 \sim \mathrm{few \; TeV} >> M_{1/2}$. In figure \ref{eoss} the allowed regions in the $m_0, m_{1/2}$ plane are displayed showing clearly the tight constraints imposed by the WMAP data in conjuction with accelerator bounds. 
%%%%%%%%%%%%%%%%%%%
%%%%%%%%%%%%%%%%%%%%%%%%
%%%%%%%%%%%%%%%%%%%%%%%%%%%%%%%%%%%%%%%%%%%%%%
In table \ref{massb}, 
upper bounds on sparticle masses are shown which are derived if the WMAP value for $\Omega_{CDM} \;h_0^2$ is imposed and the $2 \sigma$  E821 bound 
$\; 149 \times 10^{-11}<\alpha_{\mu}^{SUSY}<573 \times 10^{-11}$ is observed satisfying 
at the same time all other experimental constraints. From this table it can be seen that Supersymmetry (CMSSM) will be accessible in the LHC and to any or other linear $e^+ e^-$ collider with center of mass energy $\geq 1.1 \; TeV\;$ \cite{Lahanas:2003yz}.   
From this it is apparent the importance of the E821 results and the need of further theoretical work lifting the discrepancies concerning the muon's anomalous magnetic moment calculations. 
%%%
%%%%%%%%%%%%%%%%%%%%%%%%%%%%%%%%%%%%%%%%%%%%%%%%%%%%%%%%%%%%%%%%%%%%%%%%%%%%%%
\begin{figure}[t]
\centering
\includegraphics[width=7 cm]{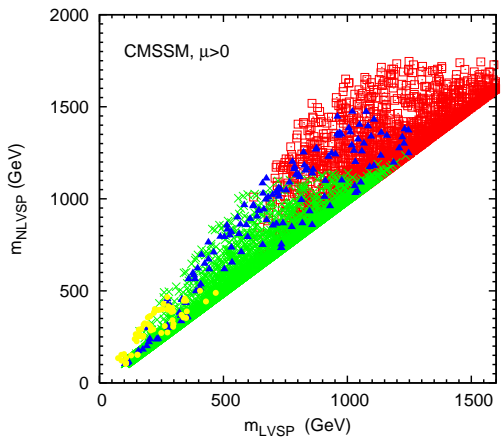} 
\caption{
Scatter plot of the masses of the lightest visible
supersymmetric particle (LVSP) and the next-to-lightest visible
supersymmetric particle (NLVSP) in the CMSSM. The darker (blue)
triangles satisfy all the laboratory, astrophysical and cosmological
constraints. For comparison, the dark (red) squares and medium-shaded
(green) crosses respect the laboratory
constraints, but not those imposed by astrophysics and cosmology.
In addition, the (green) crosses represent models which are expected to be
visible at the LHC. The
very light (yellow) points are those for which direct detection of
supersymmetric dark matter might be possible ( from \cite{Ellis:2004bx} ).}
\label{LHCscatter}
\end{figure}
%%%%%%%%%%%%%%%%%%%%%%%%%%%%%%%%%%%%%%%%%%%
%%%%%%%%%%%%%%%%%%%%%%%%%%%%%%%%%%%
\begin{table}
\centering
\caption{Upper bounds, in GeV, 
on the masses of the lightest of the neutralinos,
charginos, staus, stops and  Higgs bosons for various values of
$\tan\beta$ if the new WMAP value \cite{WMAP} for $\Omega_{CDM} h^2$ 
and the $2 \sigma$  E821 bound, 
$149 \times 10^{-11}<\alpha_{\mu}^{SUSY}<573 \times 10^{-11}$,  
is imposed ( from \cite{Lahanas:2003yz} ).
}
\label{massb}
\begin{tabular}{cccccc} 
\hline\noalign{\smallskip}
\quad $\tan\beta $ \quad & \quad ${\lsp}$ \quad & \quad  $\chi^+$ \quad 
& \quad  $\tilde{\tau} $ \quad & \quad $\tilde{t}$ \quad & \quad $Higgs$ \quad \\ 
\noalign{\smallskip} \hline\noalign{\smallskip}
%  10 \quad  &  155  \quad & 280 \quad  & 170 \quad & 580 \quad  &  116 \\
  10  &  155 & 280  & \, 170 & \, \, 580  &  116 \\
%  15 \quad  &  168 \quad  & 300  \quad & 185 \quad & 640 \quad  &  116 \\
  15 &  168  & 300  & \, 185 & \, \, 640  &  116 \\
  20 &  220  & 400  & \, 236 & \, \, 812  &  118 \\
  30 &  260  & 470  & \, 280 & \, \, 990  &  118 \\
  40 &  290  & 520  & \, 310 & \, \, 1080 &  119 \\
  50 &  305  & 553  & \, 355 & \, \, 1120 &  119 \\
  55 &  250  & 450  & \, 585 & \, \, 970  &  117 \\ 
\noalign{\smallskip} \hline\end{tabular}
\end{table}
%%%%%%%%%
Detailed studies have shown that the LHC will probe the region of the parameter space allowed by Cosmology and present accelerator data even if the data by E821 are not taken into account  
 \cite{Ellis:2004bx}  as is seen in figure~\ref{LHCscatter}. 
In a $\chi^2$ analysis performed in \cite{Baer:2003yh} it is shown that particular regions of the parameter space are favoured including the focus point region  and the fast s - channel Higgs resonance annihilation as shown in figure \ref{focusHB}.
%%%%%%%%%%%%%%%%%%%%%%%%%%
\begin{figure}[ht]
\centering 
\includegraphics[width=5.7cm]{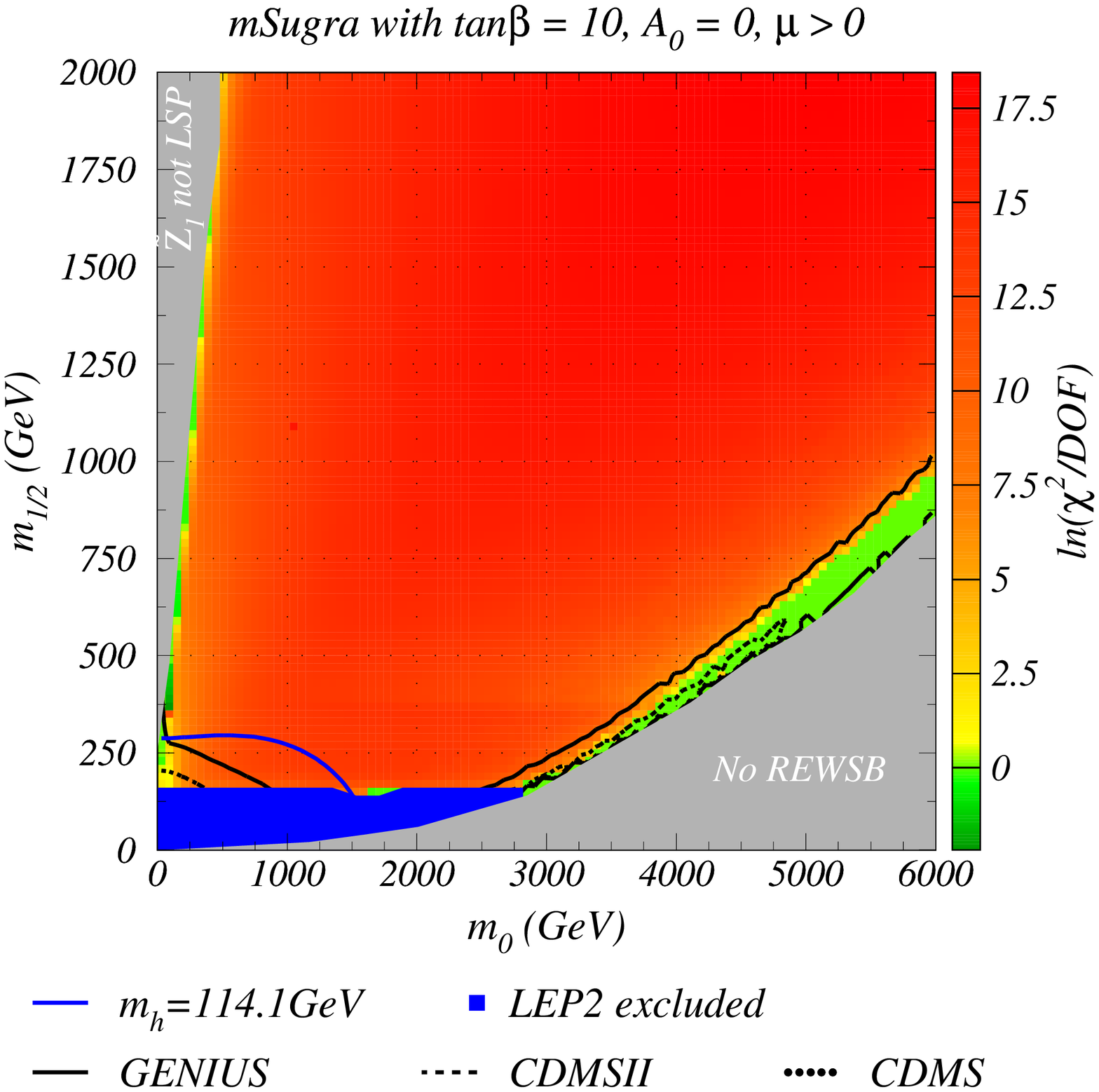} 
\hspace*{.1cm}
\includegraphics[width=5.7cm]{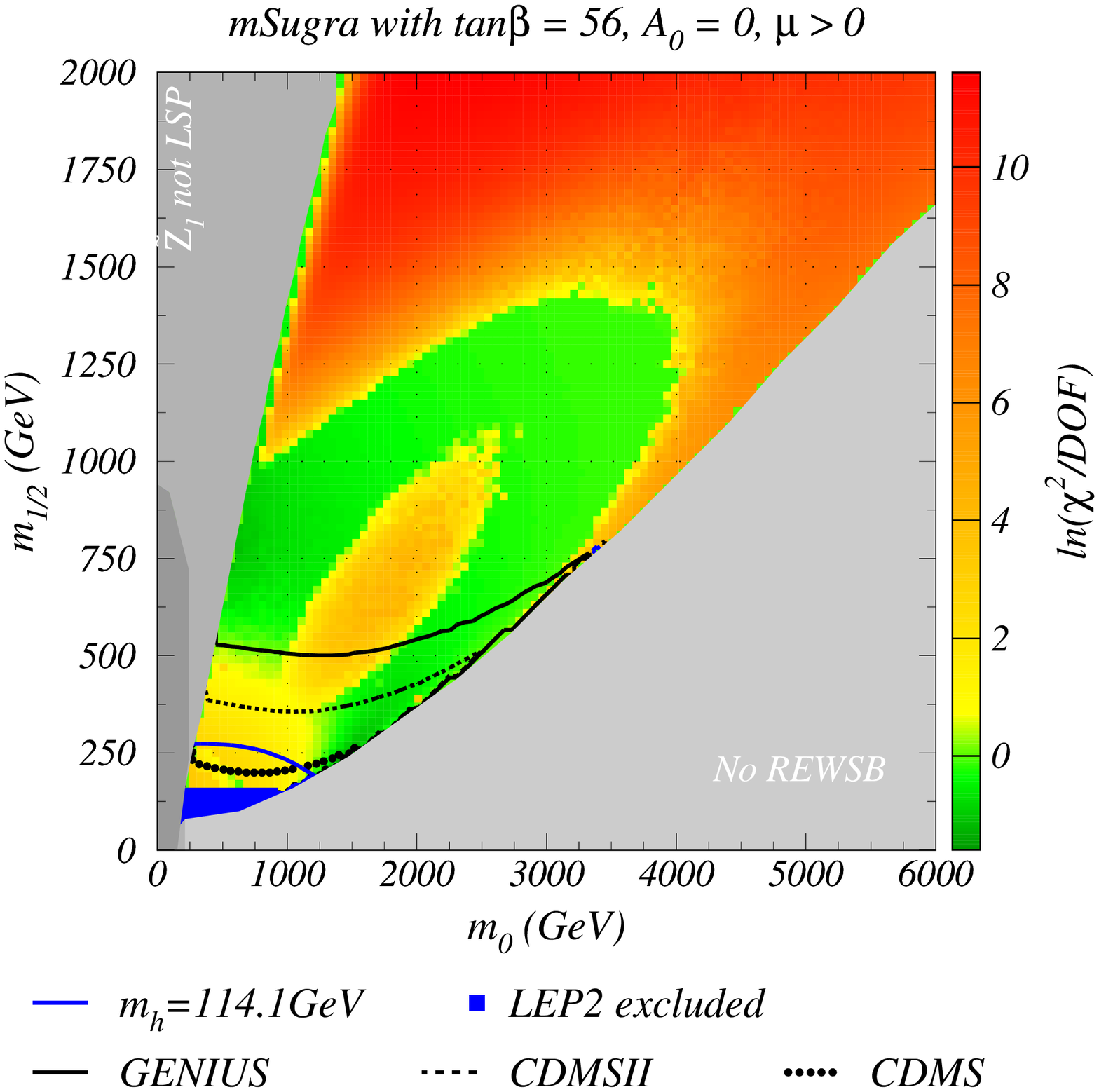} 
\caption{{
 WMAP data seem to favour ($\frac{\chi^2}{dof}<4/3$)
the HB/focus point region (moderate to large values of $\mu$, large $m_0$ scalar masses)
for almost all tan$\beta$  ( narrow stripe on the right of the Left panel - in green  ), as well as s - channel Higgs resonance annihilation ( ring-like stripe on the Right panel - in green  ) for $\mu>0$ and large tan$\beta$ ( from \cite{Baer:2003yh} ) .
}}
\label{focusHB}
\end{figure}
%%%%%%%%%%%%%%%%%%%%%%%%%%%%%% 

\begin{figure}[hb]
\centering 
\includegraphics[width=5.7cm]{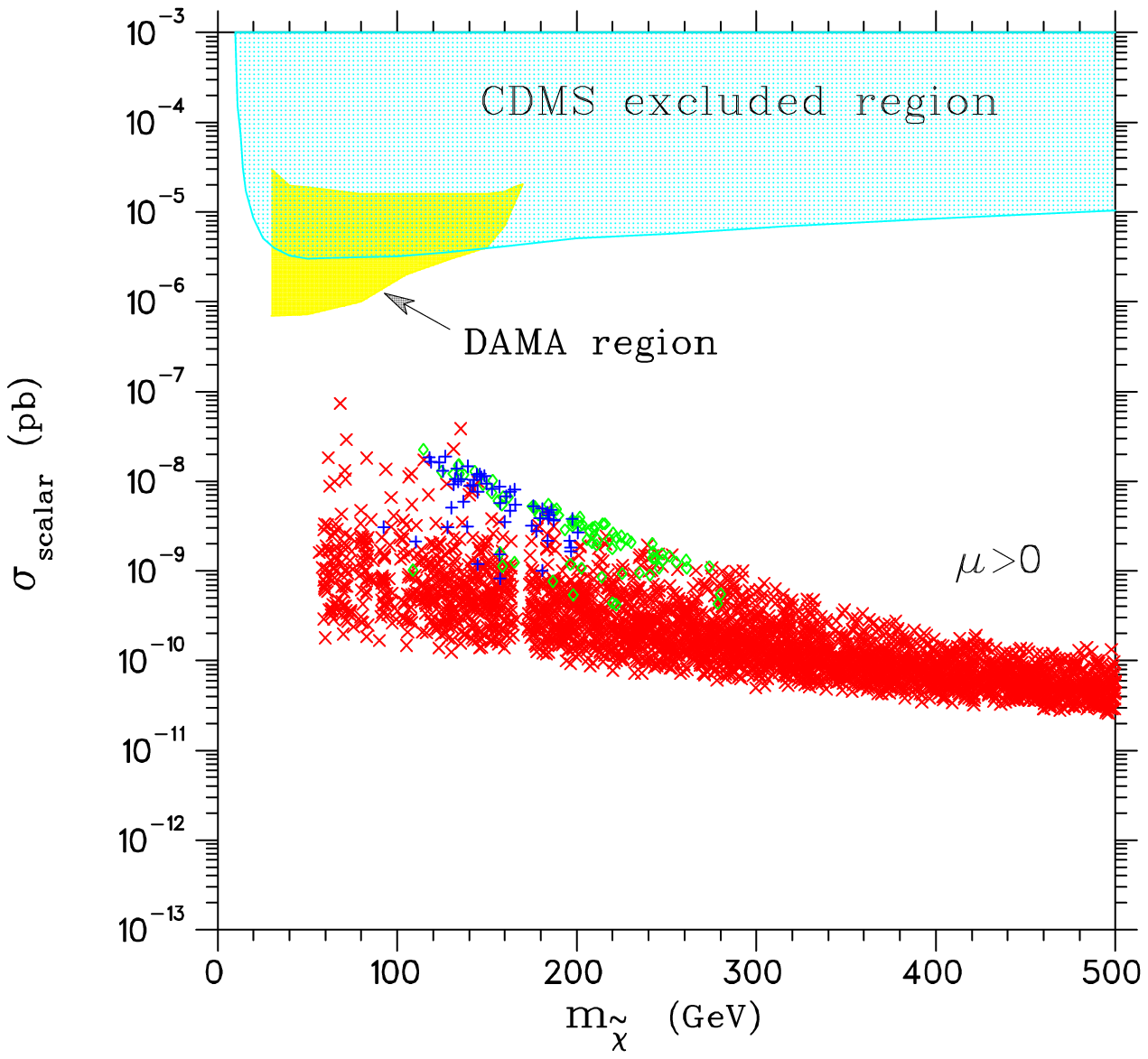} 
\hspace{0.1cm}
\includegraphics[width=5.7cm]{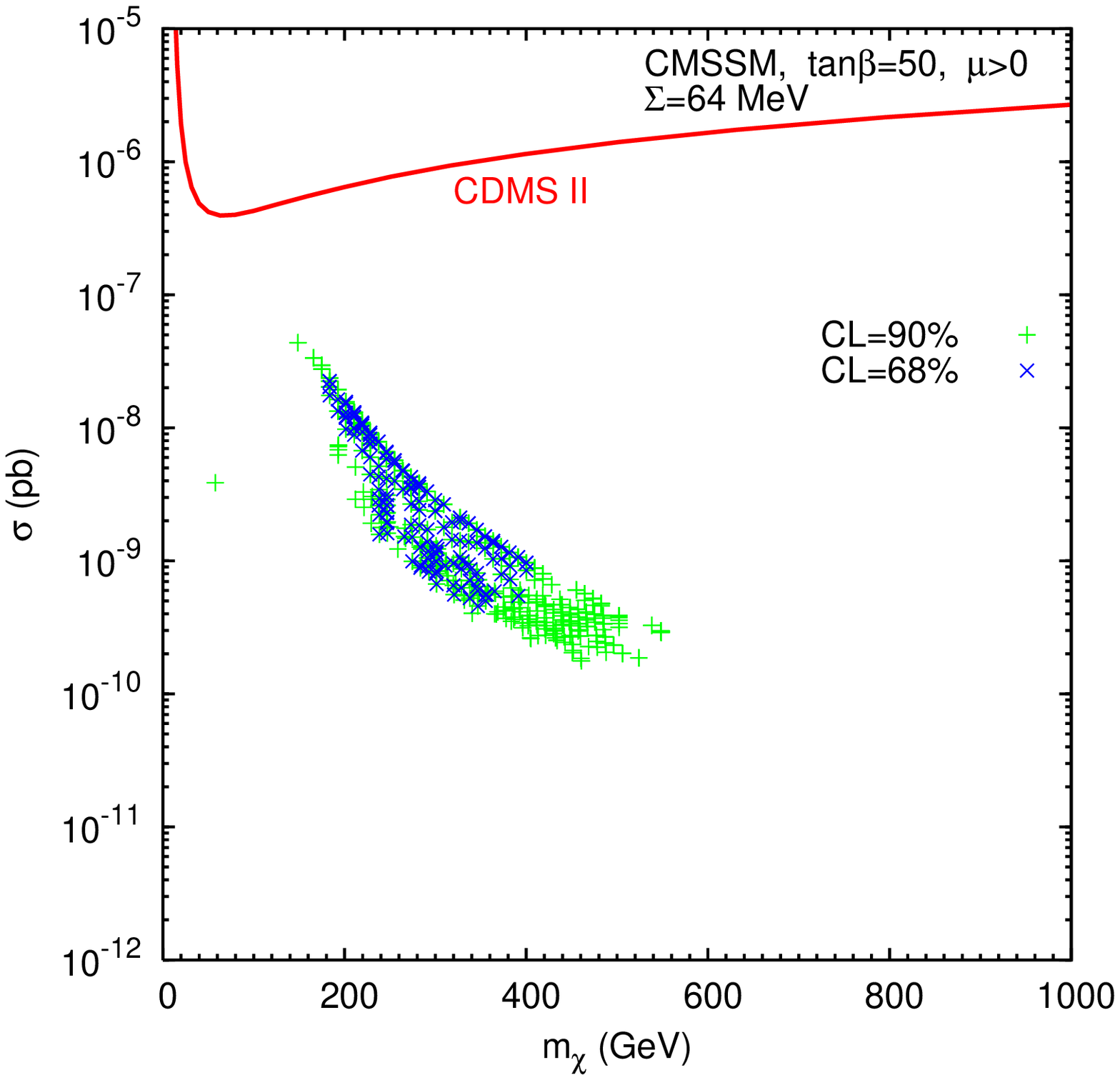} 
\caption{
 a) Left panel. Scatter plot of the spin independent neutralino-nucleon elastic  
cross section vs. $m_\chi$ predicted in the CMSSM. + signs ( in blue ) are compatible with the E821 experiment and also cosmologically acceptable. The sample consists of 40,000 random points 
( for details see \cite{Lahanas:2003yz} ). 
b) Right panel. The same as in panel a) for $\tan \beta = 50, \mu > 0$, with 
$\sigma_{\pi N} =64$~MeV. The predictions for models allowed at the 
68\% (90\%) confidence are shown by $\times$ signs - blue ($+$ signs - green) 
( from \cite{Ellis:2005mb}). 
} 
\label{fig3}
\end{figure} 
%%%%%%%%%%%%%%%%%%%%%%%%%%%%%%

\begin{figure}[t]
\centering 
\includegraphics[width=7 cm]{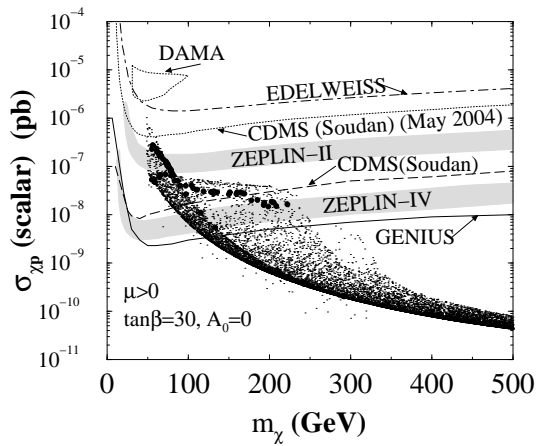} 
\vspace*{0.4mm}
\caption{
 Spin independent cross sections vs. $m_{\tilde \chi}$. The black circles are compatible  with WMAP data. The present and future DM detection limits are shown. ( from 
\cite{chatto04}). } 
%chatto
\label{scatthb2}
\end{figure}

Supersymmetric direct DM searches  through elastic neutralino-Nucleon scattering 
$\; \tilde{\chi} + N \rightarrow \tilde{\chi} + N \;$ are very important for DM detection. The rates for the spin-independent cross sections as calculated in the CMSSM are of the order 
$\; \sigma_{s.i.} \sim 10^{-7}- 10^{-8} \; \mathrm{pb} \;$ at the maximum,  still far from the sensitivity limits of CDMS II experiment as shown in figure \ref{fig3}. 
The spin independent cross section will be of interest for future  experiments, CDMS, EDELWEISS, ZEPLIN and GENIUS, which  will search for DM and will access the as yet unexplored CMSSM region \cite{directexp}. The present status and the sensitivity of future experiments, in conjuction with the theoretical predictions, is displayed in figure \ref{scatthb2}.  

The annihilations of relic particles in the Galactic halo, 
$\; \tilde{\chi} \tilde{\chi} \rightarrow \bar{p}, e^+ +... \;$, 
the Galactic center 
$\; \tilde{\chi} \tilde{\chi} \rightarrow \gamma + ...\;$
or the core of the Sun 
$\; \tilde{\chi} \tilde{\chi} \rightarrow \nu + ... \rightarrow \mu + ... \;$ 
are of great interest too. The annihilation positrons in the CMSSM seem to fall below the cosmic-ray background. The annihilated photons may be detected in GLAST experiment, while the annihilations inside the Sun may be detectable in the experiments AMANDA, ANTARES, NESTOR and IceCUBE \cite{ellis0504501}. 
%%%%%%%%%%
%%%%%%%%%%%%%%%%%%%%%%%%%%%%%%%

\subsection{ The Gravitino }

The gravitino, $g_{3/2}$, spin-3/2 superpartner of the graviton in Supergravity theories, gets non-vanishing mass after spontaneous symmetry breaking of local supersymmetry. Gravitino couples to the supersymmetric matter and if its mass $m_{\;3/2}$ is larger than that of the lightest of the neutralinos $\tilde \chi$ it can decay gravitationally to it through  
$\;g_{\;3/2}  \rightarrow \tilde{\chi} + \gamma \; $. 
If it is lighter than $\tilde \chi$  then it is the LSP and $\tilde \chi$ decays to it via  
$\; \tilde{\chi} \rightarrow g_{\;3/2} + \gamma \; $. 
Such decays produce electromagnetic radiation and 
may be upset the Big Bang Nucleosynthesis (BBN) predictions for the light element abundances. In fact the emitted Electromagnetic radiation can destroy $\;D, ^4 He, ^7 Li \;$ and/or overproduse $\; ^6 Li \;$. 

\begin{figure}[t]
\centering 
\includegraphics[width=7cm]{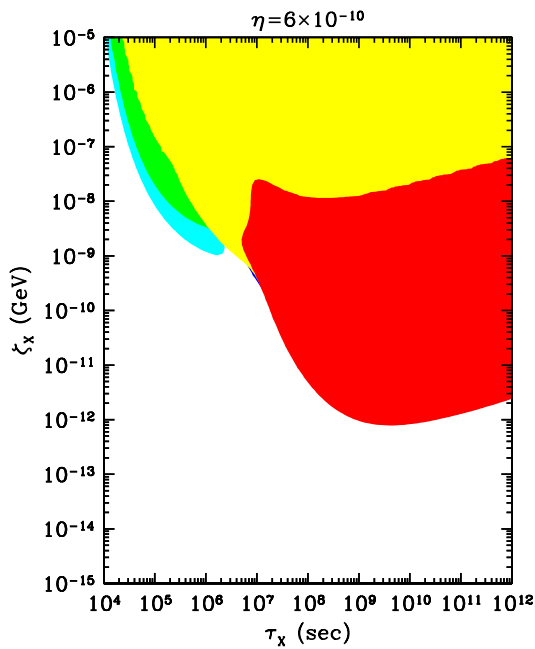} 
\caption{
Limits on $\zeta_X$, $\tau_X$ imposed by the abundances of the light elements. 
Shaded (Colored) areas are excluded . 
Dark(Red)=$^6Li$, Heavy Grey(Green)=$^7Li$,  Light Grey(Yellow)=$^6Li/^7Li$, Medium Grey(Light Blue)=$D/H $ ( from \cite{CEFO} ).   
} 
\label{zxztau}
\end{figure}

The most direct and accurate estimate of the baryon to photon ratio $\eta=n_B/n_\gamma$ is provided by the acoustic structures of the CMB perturbations \cite{WMAP} and its value, 
$\eta=6.14 \pm 0.25 \times 10^{-10} \;$, controls the BBN calculations yielding very definite predictions for the abundances of the light elements as shown in table  \ref{bbn}.  
%%%%%%%%%%%%%%%%%%%%%%%%%%%%%%%%%%%TABLE
\begin{table}
\centering
\caption{Predictions of the light element abundances}
\begin{tabular}{ccc}
\hline\noalign{\smallskip}
\hspace*{11mm}\bf{Element} & \quad \quad \quad \quad $\mathrm{\bf{Predicted}}^1 $    & \bf{Observed}  \\
\hline\hline\noalign{\smallskip}
\hspace*{11mm} $\mathrm{Y_p}$ & \hspace*{1.5cm}$0.2485 \pm 0.0005$ & \hspace*{0.3cm} $0.232 \; \mathrm{to} \; 0.258$  \\
&& \\
\hspace*{11mm} $\mathrm{D / H}$ & \hspace*{1.5cm}$2.55^{+0.21}_{-0.20} \times 10^{-5}$ & \hspace*{0.5cm} $2.78 \pm 0.29 \times 10^{-5}  $    \\
&& \\
\hspace*{11mm} $^3\mathrm{He / H}$ & \hspace*{1.5cm}$1.01 \pm 0.07\times 10^{-5}  $ & \hspace*{0.5cm} $1.5 \pm 0.5 \times 10^{-5}  $  \\
&& \\
\hspace*{11mm} $^7\mathrm{Li / H}$ & \hspace*{1.5cm}$4.26^{+0.73}_{-0.60} \times 10^{-10}$  & \hspace*{0.5cm} $1.23^{+0.68}_{-0.32} \times 10^{-10}$  \\
&& \\
\hspace*{11mm} $^6\mathrm{Li / H}$ & \hspace*{1.5cm}$1.3 \pm 0.1 \times 10^{-14}$ & \hspace*{0.5cm} $6^{+7}_{-3} \times 10^{-12}$ \\
& & \\
\hline\noalign{\smallskip}
\multicolumn{3}{l}{
$^1 \mathrm{ \bf{R. \; H.\; Cyburt, \; Phys. \; Rev. \;D  70 \; (2004)\; 023505}}$}
\end{tabular}
\label{bbn}
\end{table}
%%%%%%%%%%%%
%%%%%%%%%%%%%%%%%%%%%%%%%%%%%%%%%%%
The predicted values quoted in the table are impressively close to the observed values with the exception of the Lithium cases. In fact  the prediction for $^7\mathrm{Li}$ is predicted higher, by almost a factor of three, than the values of the astrophysical data and $^6\mathrm{Li}$ is predicted much smaller by a factor of $10^{-3}$. However these discrepancies are not disturbing due to the large sytematic errors existing in the astrophysical data. 
The decays of massive unstable products with lifetimes $\; \tau > 10^2 \; s\;$ produce Electromagnetic radiation and or hadronic showers in the Early Universe which may destroy or create nuclei spoiling the succesfull BBN predictions. The agreement with BBN imposes limits to the density of the decaying particles which depend on their lifetimes and on the value of $\eta$. These limits are shown in figure \ref{zxztau} \cite{CEFO} where   
%%%%%%%%%%%%%%%%%%%%%%%
%%%%%%%%%%%%%%%%%%%%%%%%%%%%%%%%%%%
the quantity $\zeta_X$ is defined as $\zeta_X \equiv m_X \;{n_X}/{n_\gamma}$. For instance if a decaying particle $X$ has a lifetime $\tau_X = 10^8 \; s$, the bound on the $\zeta_X$ extracted from this figure is $m_X \;n_X/n_{\gamma} < 5 \times 10^{-12} \;$, with $m_X$ in GeV units ( from \cite{CEFO}). 
These constraints were updated in \cite{Ellis:2005ii} where the effects of unstable heavy
particles were reconsidered in an attempt to reconcile the high primordial  $^7 Li$ abundance, as 
implied by the baryon-to-photon ratio, with the lower  $^7 Li$ observed in halo stars. 
%%%%%%%%%%%%%%%%%%%%%%%

In Supergravity the gravitino can either decay to $\;g_{\;3/2}  \rightarrow \tilde{\chi} + \gamma \; $ or, if it is the LSP, $\; \tilde{\chi} \rightarrow g_{\;3/2} + \gamma \; $ as mentioned earlier. These late decays can affect the light element abundances and the previously discussed limits apply. \\ \\
%%%%%%%%%%%%%%%%%%%%
{\bf a) Unstable gravitino ( $m_{3/2} > m_{\tilde \chi}$ ) :} \\
The gravitino's decay width is 
%%%%%%%%%%%%%%%%
\beq
\Gamma \;(g_{\;3/2}  \rightarrow \tilde{\chi} + \gamma ) \;=\;\frac{1}{4}\; 
\frac{m_{3/2}^{\;3}}{M_{Planck}} \; O_{\tilde{\chi} \tilde \gamma}^2 \; \; ,
\eeq
%%%%%%%%%%%%%%% 
Where $ O_{\tilde{\chi} \tilde \gamma}$ is the matrix element relating the mass eigenstate 
$\tilde{\chi}$ to the photino $\tilde \gamma$. Assuming for the sake of the argument that $\tilde{\chi}$ is a bino,  $\tilde{\chi} = \tilde{B}$, this matrix element is  $ O_{\tilde{\chi} \tilde \gamma} = \cos{ \theta_w} $. Then if no other decay is significant 
%%%%%%%%%%%%%%%%%
\beq
\tau_{3/2} \;=\; 2.9 \times 10^8 \; {\left(  \frac{100 \; GeV}{m_{3/2}} \right)}^3 \; s \;.
\eeq
%%%%%%%%%%%%%%%%%
With a gravitino mass in the range $100\; GeV - 10 \; TeV$ the gravitino's lifetime is 
$\;\tau_{3/2}= 10^2 \;-\; 10^8 \; s \;$, so it is indeed a late decaying particle. On the other hand the  thermal production of the gravitinos is estimated to be \cite{Buch}  
%%%%%%%%%%%%%%%%%%%%
\beq
Y_{3/2}\; \equiv \;\frac{n_{3/2}}{n_\gamma}\;=\;
1.2 \times 10^{-11} \; (\;1 + \frac{m_{\tilde g}^2}{12 \;m_{3/2}^2} \;) \; 
\frac{T_R}{10^{10}\;GeV} \;,
\eeq
%%%%%%%%%%%%%%%%%%%
with $T_R$ the maximum temperature reached in the Universe. With reasonable values of the gluino and gravitino masses $m_{\tilde g}, m_{3/2}$
%%%%%%%%%%%%%%%%%%%%%
\beq
Y_{3/2}\; \simeq \;
( \;0.7 \;-\;2.7 \;) \times 10^{-11} \; \frac{T_R}{10^{10}\;GeV} \;. \label{gravi1}
\eeq
%%%%%%%%%%%%%%%%%%%
From the BBN limits which we discussed previously we can infer limits on $Y_{3/2}$. For instance for a gravitino mass $ \;100\;GeV$ with lifetime $\tau_{3/2}=10^8 \;s$ we get 
%%%%%%%%%%%%%%%%%%%%%
\beq
Y_{3/2}\; \leq \;
5 \times 10^{-14}  \; , \label{gravi2}
\eeq
%%%%%%%%%%%%%%%%%%%
and relations (\ref{gravi1}) and (\ref{gravi2}) are  combined to yield an upper limit on $T_R$ 
%%%%%%%%%%%%%%%
\beq
T_R \; < \; 7 \times 10^7 \;GeV  \;. \label{limitt}
\eeq
%%%%%%%%%
\noindent
\hspace*{-2mm}
{\emph { This limit is much smaler than the reheating temperature expected in inflationary models $\; \;T_R \simeq 10^{12} \; GeV\;$. }} 
Therefore with unstable gravitinos we need an explanation to resolve this problem. \\ \\
%%%%%%%%%%%%%%
{\bf a) Stable gravitino ( $m_{3/2} < m_{\tilde \chi}$ ) :} \\
With a stable gravitino the next to it NSP ( Next Supersymmetric Particle ), which can be the $\tilde \tau$ or $\tilde \chi$, can decay to it. For the neutralino $\tilde \chi$ decay we have for instance 
%%%%%%%%%%%%%%%%
\beq
\Gamma \;(\; \tilde{\chi}  \rightarrow g_{\;3/2} + \tilde \gamma ) \;=\;\frac{1}{16 \; \pi}\; 
\frac{O_{\tilde{\chi} \gamma}^2 }{M_{Planck}^2}\;
\frac{m_{\tilde \chi}^{\;5}}{m_{3/2}^2}  \; 
{\left( \;1 - \frac{m_{3/2}^2} {m_{\tilde \chi}^{\;2}} \; \right) }^{\;3} \;
\left( \;\frac{1}{3} + \frac{m_{3/2}^2}{m_{\tilde \chi}^{\;2}} \; \right) \;
\eeq
%%%%%%%%%%%%%%% 
with $\; O_{\tilde{\chi} \tilde \gamma}= O_{1 \tilde{\chi}} \cos{\theta_w} + 
O_{2 \tilde{\chi}} \sin{\theta_w} \;$. The DM gravitino is produced via the decay 
$\; \mathrm{NSP} \rightarrow g_{\;3/2} + \gamma \; $ and its relic density is
%%%%%%%%%%%%%%%%
\beq
\Omega_{3/2}\; h_0^2\;=\;\frac{m_{3/2}}{m_{NSP}}\;\Omega_{NSP} \;h_0^2 \; 
< \;\Omega_{NSP} \;h_0^2 \;. \label{nspb}
\eeq
%%%%%%%%%%%%%%%%
From the BBN constraints and especially from $^6 Li$ abundance 
%%%%%%%%%%
\beq
\frac{n_{NSP}}{n_{\gamma}} \;<\;5 \times 10^{-14}\; \left( \frac{100 \; GeV}{m_{NSP}} \right)\;, 
\label{nspb2}
\eeq
%%%%%%%%%%
for $\tau_{NSP} = 10^8 \; s $, before the decay of the NSP. Using 
$ n_B / n_{\gamma} = 6 \times 10^{-10}$ we have the bounds 
%%%%%%%%%%%%%%%
\beq
\Omega_{NSP} \;h_0^2 \; < \; \frac{10^{-2}}{m_{NSP}} \;\Omega_B \;h_0^2 
\;<\; 10^{-2}\; \Omega_B \;h_0^2 \sim 2 \times 10^{-4} \; . \label{smallb}
\eeq
%%%%%%%%%%%%%%%%
Plugging (\ref{smallb}) into (\ref{nspb}) yields the bound 
%%%%%%%%%%%%%%%%%
\beq
\Omega_{3/2}\; h_0^2\;<\; 2 \times 10^{-4} \; !  \; \label{toosmall}
\eeq
%%%%%%%%%%%%%%%%
This is too small far away from $\sim 0.1 $ required by the cosmological data. At this point recall that values in the range $0.1$ are generic in supersymmetric models having the lightest of the neutralinos as the LSP. However in deriving this estimate we assumed a lifetime for the NSP of the order of $\sim 10^8 \;s$. For shorter lived NSPs this tight constraint is relaxed. However still  
$\; \Omega_{3/2}\; h_0^2\;< \;\Omega_{NSP} \;h_0^2 \;$ and on account of the fact that 
$\;\Omega_{NSP} \;h_0^2 \;\sim\;0.1 \;$ we may need some supplementary mechanisms to produce gravitinos as for instance reheating after inflation in addition to NSP decays. 

\begin{figure}[t]
\centering 
\includegraphics[width=5.7cm]{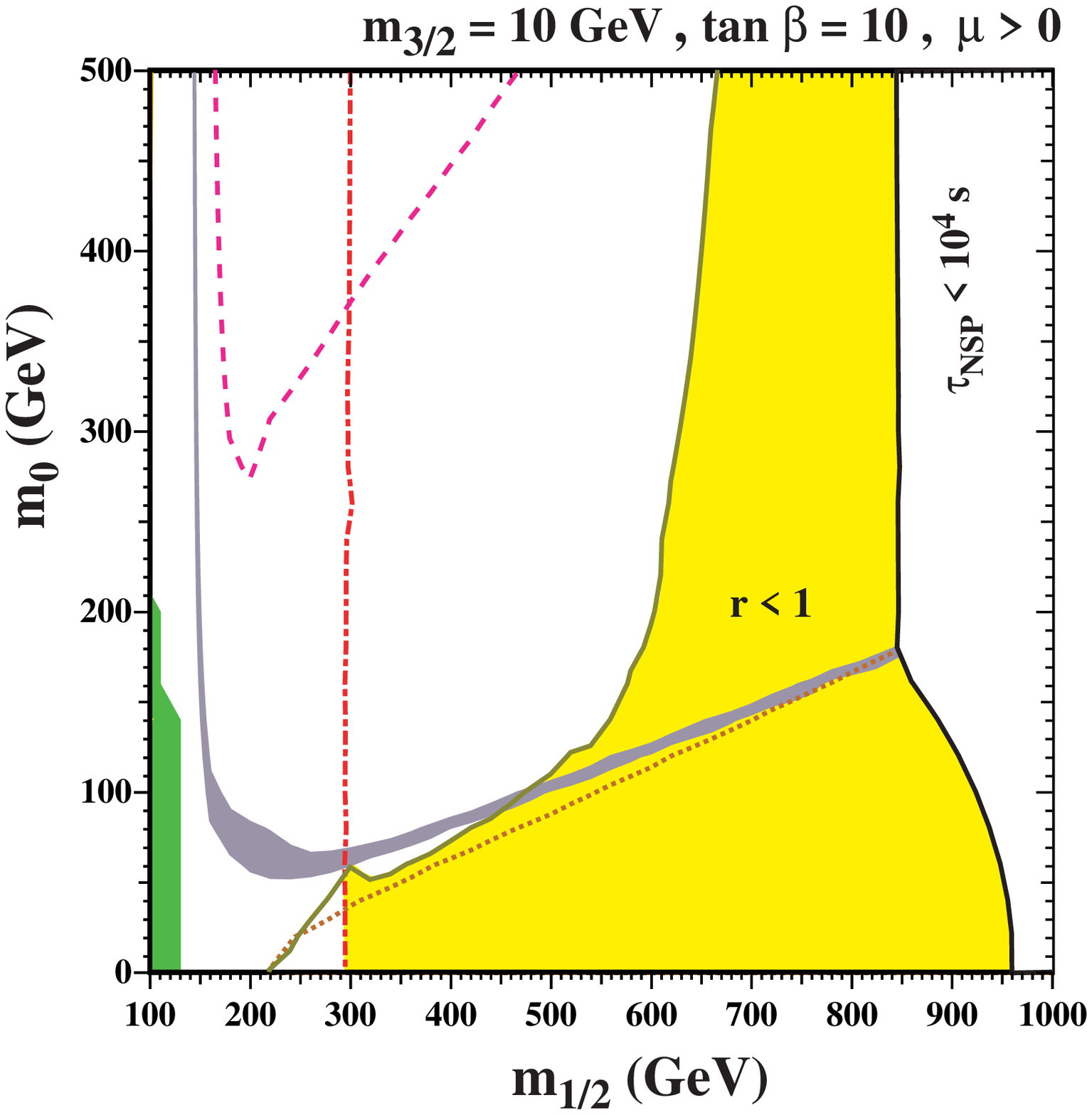} 
\hspace{0.1cm}
\includegraphics[width=5.7cm]{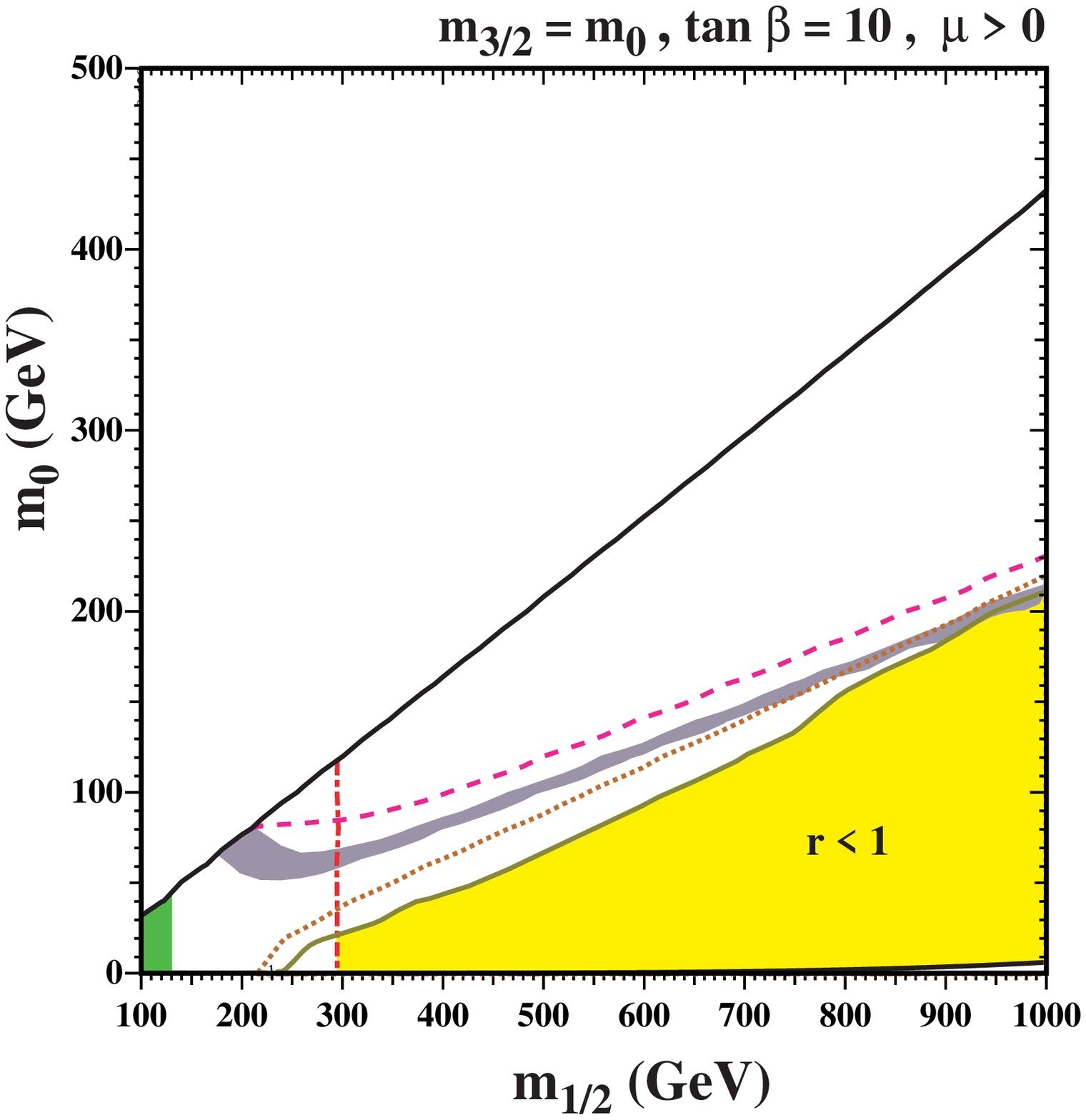} 
\caption{
The $(m_0, M_{1/2})$ planes for $\tan \beta\;=\;10$ and values of the gravitino mass equal to 
$10 \;GeV$ ( left panel ) and $m_{3/2}=m_0$ ( right panel ). In the light grey (yellow) shaded area designated by $r<1$ we have $\;\Omega_{3/2}\; h_0^2 < 0.129 \;$ and the BBN limits are observed. 
( from \cite{EOSSgrav} )
}
\label{span1}
\end{figure}

In the analysis of \cite{EOSSgrav}, whose arguments we closely followed in the previous discussion, the authors calculate the relic density the NSP would have today if it had not decayed, 
$\; \Omega_{NSP} \;h_0^2 =3.9 \times 10^7 \; \zeta_{NSP} \; GeV^{-1}$, compute the lifetime $\;\tau_{NSP}$ and impose the detailed bounds from BBN on $\; \zeta_{NSP} \;$. They delineate regions of the parameter space where the gravitino relic density $\;\Omega_{3/2}\; h_0^2\;=\;\frac{m_{3/2}}{m_{NSP}}\;\Omega_{NSP} \;h_0^2 $ is less than $0.129$ the highest CDM matter density allowed by WMAP data at $2 \sigma$. In their analysis they do not consider regions in which  
$\; \tau_{NSP} < 10^4 \; s \;$.  For such short lifetimes the hadronic decays should be considered too which would give additional constraints  strengthening  the limits on gravitino DM derived on the basis of the Electromagnetic showers only
%%%%%%%%%%
{\footnote{ 
The hadronic BBN constraints are systematically disussed in \cite{Steffen:2006hw}.
}}
.
%%%%%%%%
Some sample outputs of the limits imposed are shown in figure \ref{span1}. 
%%%%%%%%%%%%%%%%%%%%%%%%%%%%%
%%%%%%%%%%%%%%%%%%%%%%%%%%%%%%%
In the light grey (yellow) shaded areas, designated by $r<1$,  
$\;\Omega_{3/2}\; h_0^2 < 0.129 \;$ and the BBN limits, as well as all accelerator data, are observed. In the allowed domain the relic density is typically less than that favoured by astrophysics and cosmological data and as already remarked supplementary mechanisms for the production of gravitinos may be needed for these to constitute the DM of the Universe. 

\section{Conclusions}
Supersymmetric Cold Dark Matter is offered as a plausible explanation for explaining the missing mass of the Universe. Supersymmetric Theories will be tested in the near future at LHC or other high energy machines and will confirm or refute the existence of supersymmetry with consequences of paramaount importance for the future of the small scale physics. The confirmation of the existence of supersymmetric matter in the laboratory will have important consequences for Astrophysics too. The abundances of the LSP neutralinos in supersymmetric models come out to be in the right amount as required by astrophysical observations and it is the ideal candidate for explaining the DM problem. The gravitinos are also good CDM candidates but their  case needs further investigation. In the most of the parameter space describing supersymmetric theories the relic density of the gravitino is less than that required by observations and significant thermal gravitino production is needed in addition to the NSP decay mechanism. The precision of the cosmological data is a valuable input for particle physics. In conjuction with the laboratory experiments they put severe phenomenological constraints which particle physicists should observe 
enabling them to learn more on fundamental scale physics from the Universe. \\ \\
%%%%%%%%%%%%%%%%%%%%%%%%%%%%%%%%%
\noindent 
{\bf Acknowledgements} \\ 
%\noindent 
I thank the organizers for inviting me to enjoy the pleasant atmosphere of the Third Aegean School on Cosmology and my colleagues D. V. Nanopoulos and V. C. Spanos for the enjoyable collaboration we have had on the Dark Matter issues. Part of the material of this lecture is based on work done with them.
This work is co-funded by the European Social Fund (75 \%) and National Resources (25 \%) - 
EPEAEK B - PYTHAGORAS. 
%%%%%%%%%%%%%%%%%%%%%%%%%%%%%%%%%%%%%%%%  
%%%%%%%%%%%%%%%%%%%%%%%%%

\end{document}